\documentclass[numbered,pdftex]{ohio-etd}

\usepackage[square,sort&compress,numbers]{natbib} 
\usepackage{textcomp}
\usepackage{amssymb}  
\usepackage{bm}       
\usepackage{booktabs} 
\usepackage{dcolumn}  
\usepackage{multirow} 
\usepackage{graphicx} 
\usepackage[printonlyused]{acronym}
\usepackage{mathtools}
\usepackage{subcaption}
\usepackage{float}
\usepackage{ragged2e}
\usepackage{mathrsfs}
\graphicspath{{figures/}} 

\degree    {MS}              
\graduation{April}{2017}    

\title     {Micromechanical Analysis of Strength of Polymer Networks with Polydisperse Structures}
\author    {Mohammad J.}{Tehrani} 

\advisor   {Alireza Sarvestani}{Assistant Professor of Mechanical Engineering}
\dean      {Dennis Irwin}{Dean, Russ College of Engineering and Technology}

\program   {Mechanical Engineering}            
\department{Department of Mechanical Engineering} 
\college   {Russ College of Engineering and Technology}   

\abstract  {The effect of network chain distribution on mechanical behavior of elastomers is one of the long standing problems in rubber mechanics. The classical theory of rubber elasticity is built upon the assumption of entropic elasticity of networks whose constitutive strands are of uniform length. The kinetic theories for vulcanization, computer simulations, and indirect experimental measurements all indicate that the microstructure of vulcanizates is made of polymer strands with a random distribution of length. The polydispersity in strand length is expected to control the mechanical strength of rubber as the overloaded short strands break at small deformations and transfer the load to the longer strands. The purpose of the contributions presented in this thesis is to present simple theories of rubber mechanics that take into account the length distribution of strands and its effect on elasticity and the onset of bulk failure in unfilled and filled elastomers. In unfilled system, the population of short chains are identified as the culprits for damage initiation. Upon deformation of a polydisperse network, shorter strands break at considerably smaller stretches compared to the longer
ones. The network alteration continues concurrent with increasing deformation and controls the onset of mechanical failure. In the filled networks, the degradation in network mechanical behavior is assumed to be controlled by the adhesive failure of the short strands adsorbed onto the filler surface. The finite extensibility of the short adsorbed strands is a key determinant of mechanical strength.
}
\dedication{Dedicated to my wife, Yasi, for her unconditional support and love through all these years\\ also to my respectfull parents, Mina and Alireza, for their support,encoregment, and inspiration .}

\acknowledgments{ I would like to express my gratitute to my advisor Dr. Alireza Sarvestani for his supports. Without his insightful guidance, priceless advices, and endless help this journy would not be accomplished. \par
 I am also gratful to my thesis committee members, Dr. Khairul Alam, Dr. Munir Nazzal, and Dr. Arash Yavari for their advice and direction during this research.\par
At last but not least, I would like to thank my friends and colleagues, Zahra Ghalamzan, Mohammad Hossein Moshaei, and Tejas Bhave for their support.
}

 \notables  

\begin{document}

\makefrontmatter    
\justify
\chapter{Introduction}
\section{Background}
 A major assumption of classical rubber elasticity is the uniform distribution of crosslink points or the monodispersity of constitutive rubber strands. The traditional crosslinking techniques, however, are essentially uncontrolled processes and thus the ideal network structure is almost never found in practice. Betisde et al. \cite{bastide1990scattering} first predicted that the fluctuation in crosslink density in polymer gels is inherent to the network structure and will appear even if the crosslinking reaction is stopped far beyond the percolation threshold. Subsequent light and neutron scattering studies confirmed the existence of such spatial heterogeneity \cite{onuki1992scattering,bastide1988polymer}.

 The heterogeneity of crosslink density in real polymer networks can be attributed to the slowdown in dynamics of polymer chains upon crosslinking and association with other chains \cite{vilgis1994statics}. Crosslinked chains show significantly lower mobility as compared to free primary chains and thus have higher chance to further contribute to crosslinking process. This eventually leads to the formation of local clusters of short strands and polydispersity in strands length. The irregular structure of the networks formed at the gel point is preserved throughout the course of crosslinking. The clustering may also appear due to increase in localized polymerization reaction induced by peroxide radicals, autocatalytic reactions during vulcanization, or high energy radiations by free radicals \cite{van2000cross}.

 Depending on the type of crosslinking system used to form the network, the heterogeneous internal structure of elastomers may or may not feature scale invariance [4,6]. The networks that possess scale invariance (i.e., fractal structure at certain length scales) are formed by association of several percolation clusters with self-similar structure. These networks are generally formed by linking of primary chains which are functionalized at their ends. The structure and properties of such networks have been extensively studied by Vilgis and Henrich \cite{vilgis1994statics,vilgis1992new,vilgis1992rubber,vilgis1994dynamics}. The networks without self-similarity are formed by random association of long primary chains at an arbitrary monomer along the chain, similar to traditional vulcanization process.The random heterogeneity of these 
haphazardly joined chains is due to decreased mobility of the long primary chains.

 The effect of the molecular weight distribution of polymer strands on the overall mechanical behavior of networks is of great research and technological relevance. Most of the available experimental measurements concern the mechanical properties of heterogeneous networks formed by end-linking of primary crosslinkers. In an extensive series of articles, Mark and his co-workers studied the mechanical properties of these networks formed by end-linking reaction of chains with multi-modal distribution of length (see \cite{mark2003elastomers} and the references therein). The results of their comprehensive studies on bimodal networks point to a great enhancement in ultimate mechanical properties of the network. A better toughness and longer elongation at break were reported for bimodal networks. These results were attributed to the distribution of stress between the short and long chains. The enhancement in strength is primarily due to the limited deformability of non-Gaussian short chains. Following the rupture of short chains, the stress is transferred to the long strands which exhibit larger deformation at break.

\section{Mechanical Modeling of Elastomers}
 A number of phenomenological and micromechanical constitutive models are proposed to predict the mechanical behavior of the elastomers at finite deformations \cite{ali2010review,charlton1994review}. The phenomenological models generally use the invariants of deformation tensor in order to estimate the stored strain energy density function. Neo-Hookean, Mooney-Rivlin, and Ogden models are among the frequently used phenomenological models for elastomers \cite{ali2010review}. Besides the lack of physical interpretation of some parameters, these models show poor accuracy at finite deformations close to the locking stretch.\par
\begin{figure}[H]
	\centering
	\includegraphics[width=0.8\linewidth]{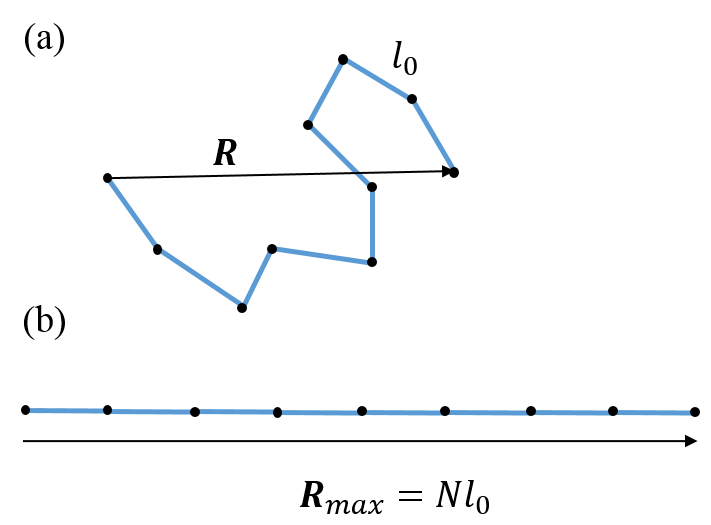}
	\caption{(a) Equilibrium configuration of a polymer strand. (b) Contour length or fully stretched polymer strand.}
	\label{fig:configurations}
\end{figure}
The micromechanical approaches to rubber elasticity are based on evaluation of entropic elasticity of a single chain followed by proper homogenization method. 
In what follows, we briefly describe the entropic elasticity of a single polymer chain. Consider a chain formed by $j$ statistical segments with end-to-end vector $\textbf{R}$ suspended in a solvent with constant temperature $T$ (Figure 1.1). The free energy of the chain can be written as
\begin{equation}
F(\textbf{R},j)=U(\textbf{R},j)-TS(\textbf{R},j)
\end{equation}

\noindent where $U$ is internal energy and $S$ shows the entropy of the chain. The internal energy of the chain is due to the interatomic interactions between the chain's constituents and as such it is independent from end-to-end distance, that is $U(\textbf{R},j)=U(j)$. The entropy of the chain can be obtained using the Boltzmann equation 

\begin{equation}
S(\textbf{R},j)=k_{B} \ln P(\textbf{R},j) +S(j)
\end{equation}
$k_{B}$ is the Boltzmann constant, $P(\textbf{R},j)$ shows the probability distribution of having a chain with $j$ statistical segments with end-to-end distance vector $\overrightarrow{R}$, and $S(j)$ is the reference entropy independent from $\textbf{R}$. For long polymer chains formed by large number of the statistical segments the probability distribution,$P(\textbf{R},j)$, is Gaussian \cite{colby2003polymer}

\begin{equation}
P(\textbf{R},j) = \bigg(\frac{3}{2 \pi j \ l^2} \bigg)^{\frac{3}{2}}  \exp \bigg(\frac{-3 \textbf{R}^2}{2  j \  l^2} \bigg)
\end{equation}

\noindent where $l$ is the length of each statistical segment. Substitution of Eq (1.3) into Eq (1.2) yields

\begin{equation}
S(\textbf{R},j) = -\frac{3}{2} \ k_{B} \ \frac{\textbf{R}^2}{j \ l^2}+S_{0}
\end{equation}

\noindent where $S_{0}$ is a reference entropy. Substitution of Eq (1.4) int Eq (1.1) provides the following expression for the chain free energy

\begin{equation}
F(\textbf{R},j)=\frac{3}{2} \ k_{B} \ \frac{\textbf{R}^2}{j \ l^2}+F_{0}
\end{equation}

\noindent where $F_{0}$ is a reference energy independent from $\textbf{R}$. \par

If the end-to-end distance $\textbf{R}$ is perturbed by $d\textbf{R}$ during a reversible isothermal process, the recoiling force $\textbf{f}$ will contribute in a mechanical work equal to $\textbf{f}.\ d\textbf{R}$.    This way
$$d \ F=\textbf{f}.\ d\textbf{R}$$ or

\begin{equation}
\textbf{f}=\frac{d \ F}{ d\textbf{R}} 
\end{equation}

\noindent substitution of Eq (1.6) into Eq (1.5) yields 

\begin{equation}
\textbf{f}= \frac{3 \  k_{B} \ T}{j \ l^2}   \  \textbf{R}
\end{equation}

\noindent Eq (1.7) shows that each chain can be envisioned as a linear (entropic) spring with stiffness $$\frac{3 \  k_{B} \ T}{j \ l^2}$$\par

As mentioned previously, the Gaussian approximation for probability density  $P(\textbf{R},j)$ is good only for large $j$. If the end-to-end distance of chain $\textbf{R}$ after deformation is close to $R_{max}$ ,the contour length of chain (Figure 1.1), then the Gaussian approximation is not valid anymore\cite{colby2003polymer}. This short coming is evident by Eq (1.7) as it allows the chain to be stretched to any length provided the enough force is applied. In reality,  the chain must stiffen up as $R $ gets close to the contour length, $R_{max}$. This issue is addressed by a different model for chain elasticity, often referred to as the Langevin chain. In this model, the entropic force developed in chain is represented as

\begin{equation}
f(R)= \frac{ \  k_{B} \ T}{l} \mathscr{L}^{-1} \bigg( \frac{R}{j \ l} \bigg)  
\end{equation}

\noindent where $\mathscr{L}^{-1}$ is the inverse Langevin function defined as

\begin{equation}
\mathscr{L}(\beta) = \coth(\beta)-\frac{1}{\beta}
\end{equation}

\section{Assessment of Polydispersity in Polymer Network}

The model for entropic elasticity of a single chain as shown in previous section, predicts the force developed in the chains a function of $j$, the number of statistical segment, in the chains. This model was the precursor for  lots of other models for the elasticity of polymer networks. In these models, each strand between two crosslink of the network is assumed to follow the entropic elasticity model as described above. Treloar \cite{jones1975properties,treloar1975physics} and Wang et al.\cite{wang1952statistical} proposed the first micromechanical models for rubber elasticity at small deformations using Gaussian statistics. Further developments were based on the Langevin statistics \cite{treloar1979non,james1943theory} and took into account the finite extensibility of the strands. Due to the complexity of the Langevin statistics, simplified geometric models were proposed for rubber internal structure  \cite{beatty2003average}, such as tetrahedral model \cite{flory1961thermodynamic}, three chain model \cite{james1947theory}, and eight-chain model \cite{arruda1993three}.  \par

All of these models are based on a simplifying assumption that all network strands are monodisperse and have equal length, This is obviously a simplifying assumption. Typically vulcanization techniques are random process in nature and formation of monodisperse network in not probable in reality. \par 

Bueche \cite{bueche} and Watson \cite{watson1953chain,watson1954chain} first proposed a simple distribution function for the length of strands in a random polymer network. Consider a network formed by vulcanization of infinitely long polymer chains. Let ${n_{j}}$ be the number of strands with ${j}$ statistical segments. If the placement of crosslink points is taken to be completely random with probability ${p}$, then the probability distribution of having a strand with ${j}$ statistical segments, $P(j)$, can be expressed as

\begin{equation} 
	P(j)=\frac{n_{j}}{\sum_{j}^{}n_{j}}=(1-p)^{j-1} p
\end{equation}

\noindent where ${\sum_{j}^{}n_{j}}$ is the total number of existing strands. The assumption of completely random placement of crosslinks warrants the probability ${p}$ to be a constant and equal to the reciprocal of average strand length $\overline{j}=\frac{1}{p}$. At the limit of large $\overline{j}$ values, Eq. (1.10) leads to a distribution function

\begin{equation} \label{eq:LeghGe}
	P(j)=\frac{1}{m}\Big(1+\frac{1}{m}\Big)^{-j}
\end{equation}

\noindent where $m=\frac{1}{p}-1$. For large $m$ values, the approximation $\Big(1+\frac{1}{m}\Big)^{m}\approx e$ holds and thus the distribution (1.11) accept a simple exponential form

\begin{equation}
P(j) =\frac{1}{\overline{j}} \exp \ \big(\frac{-j}{\overline{j}}\big)
\end{equation}

Example of this exponential probability density are shown in Figure (1.2). An important feature of Eq (1.10) is a sharp decrease in the distribution of strands including a large population of short strands. This is expected to affect multiple aspects of mechanical behavior of the networks. During deformation, shorter strands may reach maximum elongation first and eventually snap due to the large entropic tension. The rupture of short and highly stretched strands continues and eventually determines the mechanical strength of the network. Assessment of the validity of this scenario is one of the main goal of this thesis.\par

 \begin{figure}[H]
 	\centering
 	\includegraphics[width=0.8\linewidth]{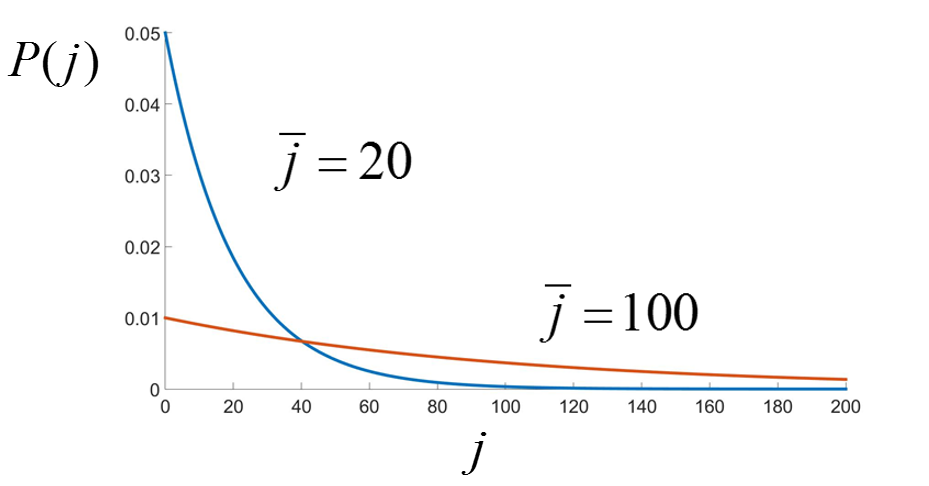}
 	\caption{Chain length distribution of strands in a polymeric network .}
 	\label{fig:configurations}
 \end{figure}

\section{Damage in Polymer Networks}

  When a network is subjected to finite deformation, its strands deform and stretch with the macroscopic deformation. Since a strands is unable to stretch beyond its contour length, at some point it has to break and become elastically inactive. Surprisingly, such a simple picture of damage initiation and failure in polymer networks has not been properly incorporated into the available constitutive models for rubber elasticity. Boyce and co-workers \cite{arruda1993three,bergstrom2005molecular} simply considered the locking stretch as the failure mode of elastomers. This is obviously an over simplification. The fracture mechanics and fatigue behavior of rubbers and rubber-like materials are manifested by evolution of microstructure of rubber and degradation of it properties at multiple spatial and time scales. Energy limiter approach \cite{trapper2008cracks,volokh2013review,volokh2010modeling,trapper2010modeling} is another model recently developed for the fracture in rubbers (Figure 1.3). As comes from the name, in this approach the strain energy is enforecd to have a saturation value, indicating the onset of damage initiation which ultimately leads to catastrophic failure The functional forms of the so-called energy barriers, however, appear to be completely arbitrary with no connection to the microstructural details of polymer network. Lumped energetic-entropic model is a more recent and probably a more realistic micromechanical approach to rubber fracture \cite{dal2009micro}. It entails some aspects of micro-, meso-, and macroscopic details of rubber structure and its evolution during loading and provides a successful predict of rubber viscoelastic behavior and durability properties. (Figure 1.4). \par
\begin{figure}[H]
	\centering
	\includegraphics[width=0.8\linewidth]{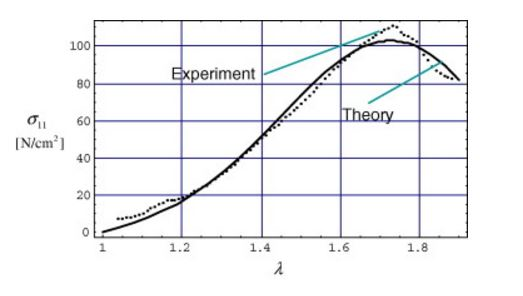}
	\caption{Cauchy stress vs. stretch in uniaxial tensile test predicted by \cite{bergstrom2005molecular} in comparison with experimental data (adapted from \cite{trapper2008cracks}).}
	\label{fig:configurations}
\end{figure}
\begin{figure}[H]
	\centering
	\includegraphics[width=0.8\linewidth]{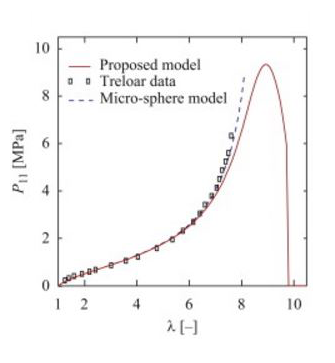}
	\caption{The first Piola-Kirchhoff stress vs. stretch predicted by micro-sphere model\cite{miehe2014phase}, the modified micro-sphere model \cite{dal2009micro} and the experimental data of Treloar \cite{treloar1979non} for  uniaxial tension tests (adapted from \cite{dal2009micro}).}
	\label{fig:configurations}
\end{figure}
\begin{figure}[H]
	\centering
	\includegraphics[width=0.8\linewidth]{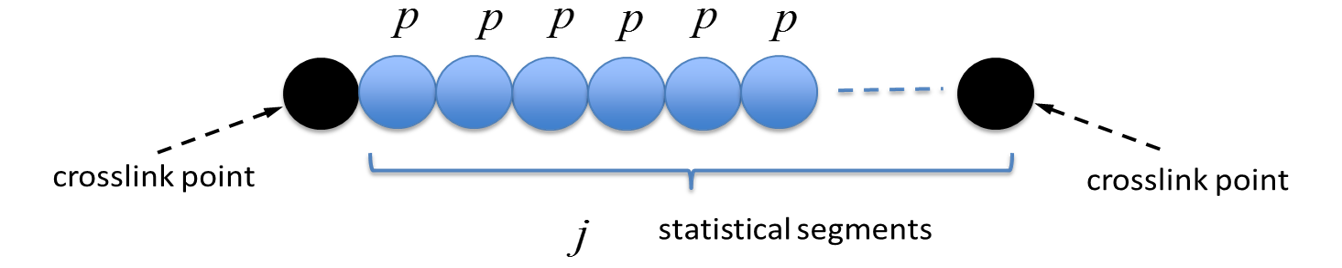}
	\caption{A polymer strand between two crosslink points. All statistical segments have the same chance, $p$, to contribute in vulcanization.}
	\label{fig:configurations}
\end{figure}

A simple model of a polymer chain can be represented by a series of statistical segments that are attached by a binding potential (Figure 1.5 ). Morse potential is a binding potential that has been frequently used  to model breakable bonds subjected to tension this potential has the general form of (Figure 1.6)

\begin{equation} \label{eq:LeghGe}
U_{M}(r)=U_{0}\Big(1-e^{-\alpha(r-r_{0})}\Big)^2
\end{equation}

where $U_{0}$ is the dissociation energy, $\alpha$ is a constant, $r$ and $r_{0}$ are the length and equilibrium length of a bond, respectively.  The anharmonic nature of the Morse potential predicts a retractive force of 

\begin{equation}
f_{e}(r)=\frac{\partial U_{M}(r)}{\partial r}
\end{equation}

and this force is maximum at 

\begin{equation}
r-r_{0}=\alpha \ln 2
\end{equation}

A single Morse bond ruptures when the retractive force exceeds 

\begin{equation} \label{eq:LeghGe}
\Big(f_{M}\Big)_{max}=\frac{\alpha U_{0}}{2}
\end{equation}

This expression is used as a single criterion for strand rupture in this thesis.
\begin{figure}[H]
	\centering
	\includegraphics[width=0.8\linewidth]{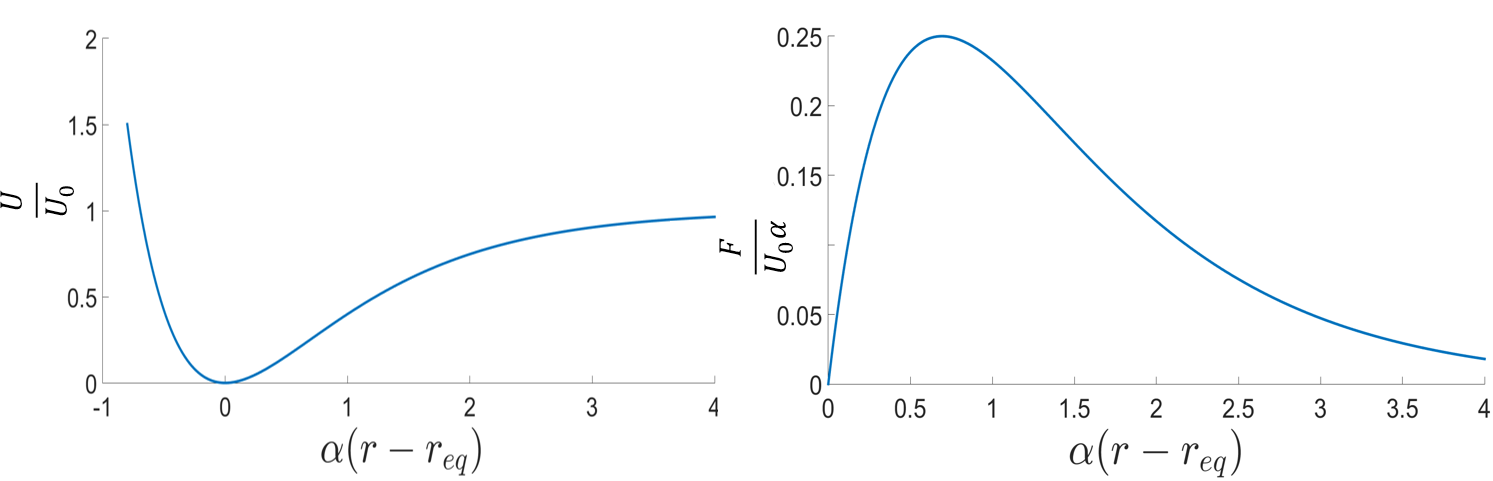}
	\caption{The variation of Morse potential and the corresponding force with the distance between two interacting particles.}
	\label{fig:configurations}
\end{figure}

\section{Effect of Filler Particles}

Most commercial rubbers are compounded with inorganic or organic fillers to obtain a better mechanical properties. Although reinforcing can not be unambiguously defined but most active fillers can improve certain properties such as stiffness,strength, abrasion and scuff resistance. Silica and carbon black are the most commonly used filler used in polymer industry.\par

Carbon black is the carbon in a colloidal form, typically obtained by combustion of petroleum products. The size of carbon black particles used in industry ranges between a few tens to a few hundreds of nanometers. Fumed silica is the particulate form of silica dioxide obtained by oxidation of silica tetrachloride. Both carbon black and silica primary particles tend to aggregate and form agglomerates clusters with fractal structure. When compounded with rubber, the strong surface interaction leads to formation of a polymer layer strongly adsorbed to the filler surface. Adsorption of polymer chains is driven either by disorder-induced localization of polymers to the graphite sheets at the carbon black surface or by strong hydrogen bonds to the hydroxyl groups on the silica particles.      
\begin{figure}[H]
	\centering
	\includegraphics[width=0.85\linewidth]{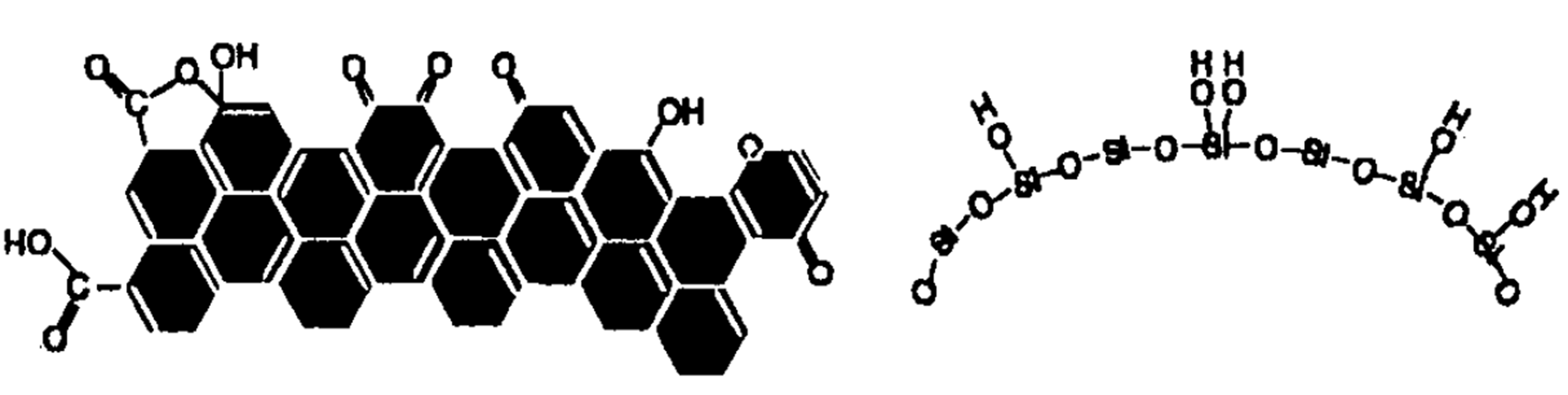}
	\caption{Surface chemistry of carbon blacks and silicas \ \cite{wolff1994silica}.}
\end{figure}

\section{Thesis Aims}
 The ultimate goal of this research is to consider the effect of random and polydisperse structure of rubber vulcanization on rubber nonlinear elasticity and damage initiation in networks. As mentioned at the beginning of this Chapter, the internal structure of polymer networks has a great degree of randomness and heterogeneity. In reality, the strand length distribution of networks is continuous, which may range from very short to very long strands. The effect of polydispersity in strands length is essentially disregarded in available micromechanical models. This research, aims to propose a micromechanical model for the elasticity and damage initiation in elastomers with a distribution of strand length. To fulfill this aim, the effect of polydispersity parameters on the ultimate static mechanical properties of respectively unfilled and filled polydisperse polymeric network at finite deformation has been studied. We only focus on the random polydisperse networks without any spatial heterogeneity. Chapters 2 and 3 contained within the thesis document serve as prepublication manuscripts. These manuscripts have been formatted to meet the guidelines set forth by Thesis and Dissertation Services at Ohio University.

\chapter[Effect of Chain Length Distribution on Mechanical Behavior of Polymeric Networks]{Effect of Chain Length Distribution on Mechanical Behavior of Polymeric Networks\footnote{Adapted from: M. Tehrani, A. Sarvestani, Effect of Chain Length Distribution on Mechanical Behavior of Polymeric Networks, European Polymer Journal 87(2017) 136-146.\cite{tehrani2017effect} Advance online publication . DOI:http://dx.doi.org/10.1016/j.eurpolymj.2016.12.017. This manuscript version is made available under the CC-BY-NC-ND 4.0 license,  http://creativecommons.org/licenses/by-nc-nd/4.0/}}

\section{Abstract}
	The effect of network chain distribution on mechanical behavior of elastomers is one of the long standing problems in rubber mechanics. The classical theory of rubber elasticity is built upon the assumption of entropic elasticity of networks whose constitutive strands are of uniform length. The kinetic theories for vulcanization, computer simulations, and indirect experimental measurements all indicate that the microstructure of vulcanizates is made of polymer strands with a random distribution of length. The polydispersity in strand length is expected to control the mechanical strength of rubber as the overloaded short strands break at small deformations and transfer the load to the longer strands. The purpose of this contribution is to present a simple theory of rubber mechanics which takes into account the length distribution of strands and its effect on the onset of bulk failure.

\section{Introduction}
 A major assumption of classical rubber elasticity is the monodispersity of the constitutive strands, the sub-chains between the two consecutive crosslink points. The conventional crosslinking techniques, however, are essentially uncontrolled processes and hence, the formation of ideal monodisperse networks is not probable. Direct measurement of randomness in internal structure of rubber compounds is unfeasible due to insolubility of the polymer networks. First efforts to indirectly quantify the structural polydispersity of vulcanizates go back to the pioneering works of Toboslky \cite{tobolsky}, Beuche \cite{bueche}, Gehman \cite{gehmanmolecular}, and Watson \cite{berry}, using relaxometry of stressed networks or measurement of swelling pressure.

\bigskip
 The effect of  strand polydispersity on the overall mechanical behavior of polymer networks is of great research and technological importance. The simplest polydisperse networks can be formed by end-linking of functionally terminated crosslinkers with a multimodal length distribution. Mark and his co-workers conducted a comprehensive study on bimodal polymer networks (\cite{mark} and the references therein). Their results point to a great enhancement in ultimate mechanical properties of the network, namely an increase in the toughness and larger elongation at break. These findings were attributed to the distribution of stress between the short and long chains. The enhancement in strength is primarily due to the limited deformability of non-Gaussian short chains. Following the rupture of short chains, the stress is transferred to the long strands which exhibit larger deformation at break.

\bigskip
 In vulcanizates, the strand length distribution is expected to be non-uniform and range from very short to very long strands \cite{gehmannetwork}. This assumption is validated by a number of computer simulation studies. Grest and Kremer \cite{grestkremer}, for example, simulated the equilibrium structure of randomly crosslinked networks with the number of crosslinks  well above the percolation threshold. The network was formed by instantaneous crosslinking of long primary chains in a melt state. In the ideal case of completely random crosslinking, the association of chains can be regarded as statistically independent events. Theoretically, this means that the distribution of crosslink points along the primary chains must be Gaussian and hence, the distribution of strand length between crosslink points must follow a simple exponential form with a decay length. The simulation results of Grest and Kremer support these predictions.

\bigskip
 Modeling fracture and mechanical failure of polymer networks continues to be a subject of ongoing research \cite{volokh2010modeling,trapper2008cracks,dal2009micro,miehe2014phase}. The ultimate mechanical properties of polymer networks are affected by a host of influences, ranging in a wide spectrum of length scales. This includes the microstructure of a single polymer chain (e.g., helicity, interatomic potentials, crosslinking, isomerization, etc.) as well as the chain's local environment (entanglements, cracks, etc.). The focus of current study is to develop a theoretical model to evaluate the role of strand polydispersity in the bulk failure of polymer networks.\footnotemark 
\footnotetext{Following Volokh \cite{volokh2010modeling}, here, the concept of bulk failure refers to the \textquotedblleft continuum damage mechanics\textquotedblright in which the material failure is controlled by damage accumulation and evolution of internal structure of the bulk material. This approach is different from the so called \textquotedblleft cohesive zone\textquotedblright models in which the properties of bulk material remain unchanged and fracture is presented by introducing interface cohesive elements whose behavior is controlled by some traction-separation laws.}
The importance of strand length distribution for the mechanical strength of vulcanizates was first highlighted by Gehman \cite{gehmannetwork}. He proposed that upon deformation of a random network, shorter strands break at considerably smaller deformations compared to the longer ones. This deformation-induced network alteration continues concurrent with increasing deformation and controls the onset of mechanical failure. This proposition is adopted here and forms the basis of the proposed micromechanical model for the elasticity and damage initiation in elastomers with a random distribution of strand length. This study is inspired by the recent work of Itskov and Knyazeva \cite{itskovrubber} who proposed a model for rubber elasticity based on the chain length statistics. Here, their approach is advanced by introducing a failure criterion based on the interatomic pair potential and considering damage accumulation using a simple first-order kinetic theory.

\section{Model}
 Bueche \cite{bueche} and Watson \cite{watson1953chain,watson1954chain} originally proposed an expression for the strand length distribution function in a random network. Consider a network formed by vulcanization of infinitely long polymer chains. Let ${n_{j}}$ be the number of strands with ${j}$ statistical segments. If the placement of crosslink points is taken to be completely random with probability ${p}$, then the probability distribution of having a strand with ${j}$ statistical segments, $P(j)$, can be expressed as

\begin{equation} 
P(j)=\frac{n_{j}}{\sum_{j}^{}n_{j}}=(1-p)^{j-1} p
\end{equation}

\noindent where ${\sum_{j}^{}n_{j}}$ is the total number of existing strands. The assumption of completely random placement of crosslinks warrants the probability ${p}$ to be a constant and equal to the reciprocal of average strand length $\overline{j}=\frac{1}{p}$. At the limit of large $\overline{j}$ values, Eq. (2.1) leads to a distribution function

\begin{equation} \label{eq:LeghGe}
P(j)=\frac{1}{m}\Big(1+\frac{1}{m}\Big)^{-j}
\end{equation}

\noindent where $m=\frac{1}{p}-1$. For large $m$ values, the approximation $\Big(1+\frac{1}{m}\Big)^{m}\approx e$ holds and thus the distribution (2.2) accept a simple exponential form

\begin{equation} 
P(j)=\frac{1}{\overline{j}}e^{-j/\overline{j}}
\end{equation}

 Note that Eq. (2.3) represents the probability distribution of strand length in an ideally random crosslinked network, where the positions of crosslinks are taken to be statistically independent.\par

\bigskip
 Now consider a network of crosslinked flexible strands subjected to a quasi-static finite deformation. To keep the formulation simple, throughout this paper it is assumed that the network is incompressible, although extension
of the presented theory to the compressible networks is
possible. The end-to-end vector of each strand in the reference and current configurations is represented by $\textbf{R}_{0}$ and $\textbf{R}$, respectively (Figure 2.1). The network is formed by random crosslinking of the strands whose length follow distribution (2.3). The effects of other structural properties such as crystallinity or entanglement are not taken into account. 

\begin{figure}[H]
	
	\includegraphics[width=\linewidth]{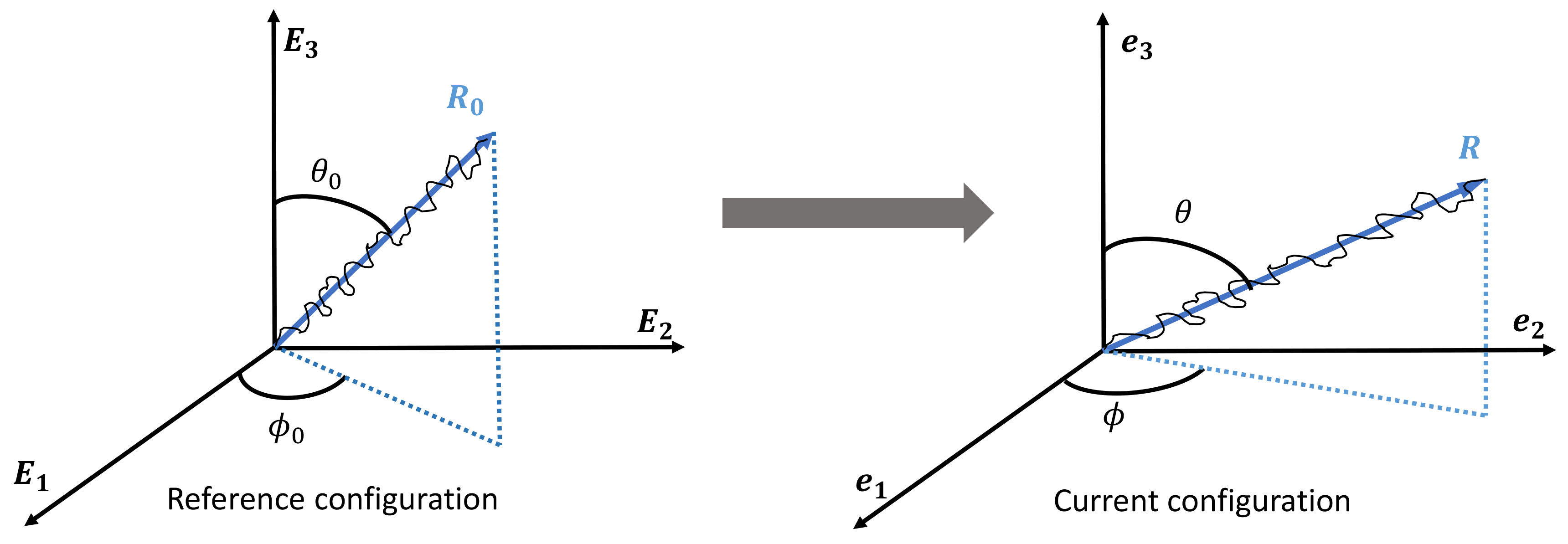}
	\caption{A network strand in the reference and current configurations.}
	\label{fig:configurations}
\end{figure}
The conformational entropy of a strand with $j$ statistical segments, stretched by $\lambda$, is

\begin{equation} 
S(\lambda,j)=-jk_{B}\bigg (\frac{\lambda}{\sqrt{j}}\beta+\ln\frac{\beta}{\sinh\beta}\bigg )-S_{0}
\end{equation}

\noindent where 

\begin{equation} 
\beta=\pounds^{-1}\bigg (\frac{\lambda}{\sqrt{j}}\bigg ) \quad
\end{equation}

 Here $k_{B}$ is the Boltzmann constant, $S_{0}$ is the reference conformational entropy, and $\pounds^{-1}$ stands for the inverse Langevin function. This way, the free energy of each strand can be written as

\begin{equation} 
w(\lambda,j)=U-TS(\lambda,j)
\end{equation}

\noindent where $T$ is the absolute temperature and $U$ is the internal energy controlled by the interatomic interactions. Following a classical approach in rubber elasticity, here the contribution of internal energy in the free energy landscape is ignored. The rupture of strands, however, is essentially controlled by the nature of this interatomic potential, as described later. Thus \cite{arruda1993three}

\begin{equation} 
w(\lambda,j)=jk_{B}T\bigg (\frac{\lambda}{\sqrt{j}}\beta+\ln\frac{\beta}{\sinh\beta} \bigg)+w_{0}
\end{equation}

\noindent where $w_{0}$ represents the deformation-independent part of the free energy.

\bigskip
 To obtain the free energy density function for a network of strands, the so called chain orientation distribution function $C(\theta,\phi)$ is used. It represents the probability distribution of having a strand with end-to-end vector $\textbf{R}$ at spherical coordinates $\theta$ and $\phi$ in the current configuration. Hence, \cite{wu1993improved}

\begin{equation} 
\int\limits_{0}^{\pi}\int\limits_{0}^{2\pi}   C(\theta,\phi) \ \sin\theta \ d\theta \ d\phi=1 
\end{equation}

\noindent The free energy density function of an ensemble of deformed strands with polydisperse length, occupying volume $\textit{V}$, can be obtained as

\begin{equation} 
W(\lambda)=\frac{\sum n_{j}}{\textit{V}}\int\limits_{0}^{2\pi}\int\limits_{0}^{\pi}\int\limits_{1}^{\infty} \ P(j) \ w \big (\lambda(\theta,\phi),j \big ) \ C(\theta,\phi) \  \sin\theta \ dj \ d\theta \ d\phi
\end{equation}

\noindent A similar function, $C_{0}(\theta_{0},\phi_{0})$,  can be defined for the orientation of strands at the reference configuration. Assuming that the strands orientation is initially random, this probability distribution can be characterized by $C_{0}(\theta_{0},\phi_{0})=\frac{1}{4\pi}$. It thus follows that \cite{wu1993improved}

\begin{equation}
C(\theta,\phi)=C_{0}\frac{sin\theta_{0}}{sin\theta}J^{-1}
\end{equation}

\noindent where $J$ is the Jacobian of deformation gradient. After substitution of (2.10) into (2.9) and taking advantage of incompressibility condition, one can obtain

\begin{equation} 
W(\lambda)=\frac{\sum n_{j}}{4\pi\textit{V}}\int\limits_{0}^{2\pi}\int\limits_{0}^{\pi}\int\limits_{1}^{\infty} \ P(j) \  w(\lambda,j) \ \sin\theta_{0} \ dj \ d\theta_{0} \ d\phi_{0}
\end{equation}

\noindent The stretch along an arbitrary direction can be expressed in the reference configuration and in terms of the macroscopic principal stretches, $\lambda_{i}$, as

\begin{equation}\lambda^{2} (\theta_{0},\phi_{0})=(\lambda_{1}  \sin\theta_{0}  \cos\phi_{0})^{2}+(\lambda_{2}  \sin\theta_{0}   \sin\phi_{0})^{2}+(\lambda_{3}  \cos\theta_{0})^{2}
\end{equation}

\bigskip
 Eq. (2.11) represents the elastic energy of a network with polydisperse strands where all strands are assumed to be elastically active. The free energy function presented by Eq. (2.7) accounts for the finite extensibility of flexible strands and diverges as the stretch approaches the ultimate locking value of $\lambda_{lock}=\sqrt{j}$ \cite{arruda1993three}. Assuming that crosslink points move in an affine fashion, shorter strands are expected to experience a larger entropic tension. As proposed by Itskov and Knyazeva \cite{itskovrubber}, the highly extended strands snap at some finite stretch and become elastically inactive. Therefore, at each direction, strands shorter than a certain length break and do not contribute to the energy function (2.11). The ultimate strength of an elastically active strand is determined either by scission of bonds along the backbone or cleavage of a crosslink. The activation energy for rupture is directly related to the nature of interatomic potential or the dissociation energy of crosslink coagents. Different harmonic and anharmonic potential functions have been used to present the energy landscape of interatomic dissociation in polymer chains \cite{doerr1994breaking,oliveira1994breaking,crist1984polymer,dal2009micro}. The Morse potential, for example, is used to predict the stiffness of a covalent bond in a nan-Gaussian polymer chain during cleavage \cite{garnier2000covalent}. Here, a Morse pair-potential is used to describe the energy barrier of debonding. That is

\begin{equation} \label{eq:LeghGe}
U(r)=U_{0}\Big(1-e^{-\alpha(r-r_{0})}\Big)^2
\end{equation}

\noindent where $U_{0}$ is the dissociation energy, $\alpha$ is a constant that determines bonds elasticity, and $r$ and $r_{0}$ show the deformed and undeformed (equilibrium) length of a bond, respectively (Figure 2.2). The strand rupture occurs when the applied force exceeds the critical value of

\begin{equation} \label{eq:LeghGe}
\Big(f_{M}\Big)_{max}=\frac{\alpha U_{0}}{2}
\end{equation}

\noindent beyond which the bonds become unstable. Eq. (2.14) limits the maximum force that can be developed in each strand. It is assumed that the bond cleavage occurs when this force equals the restoring entropic force between the crosslinks, $f_{e}$, defined as

\begin{equation}
f_{e}=\frac{\partial w}{\partial R}
\end{equation}

\begin{figure}
	\centering
	\includegraphics[width=0.75\linewidth]{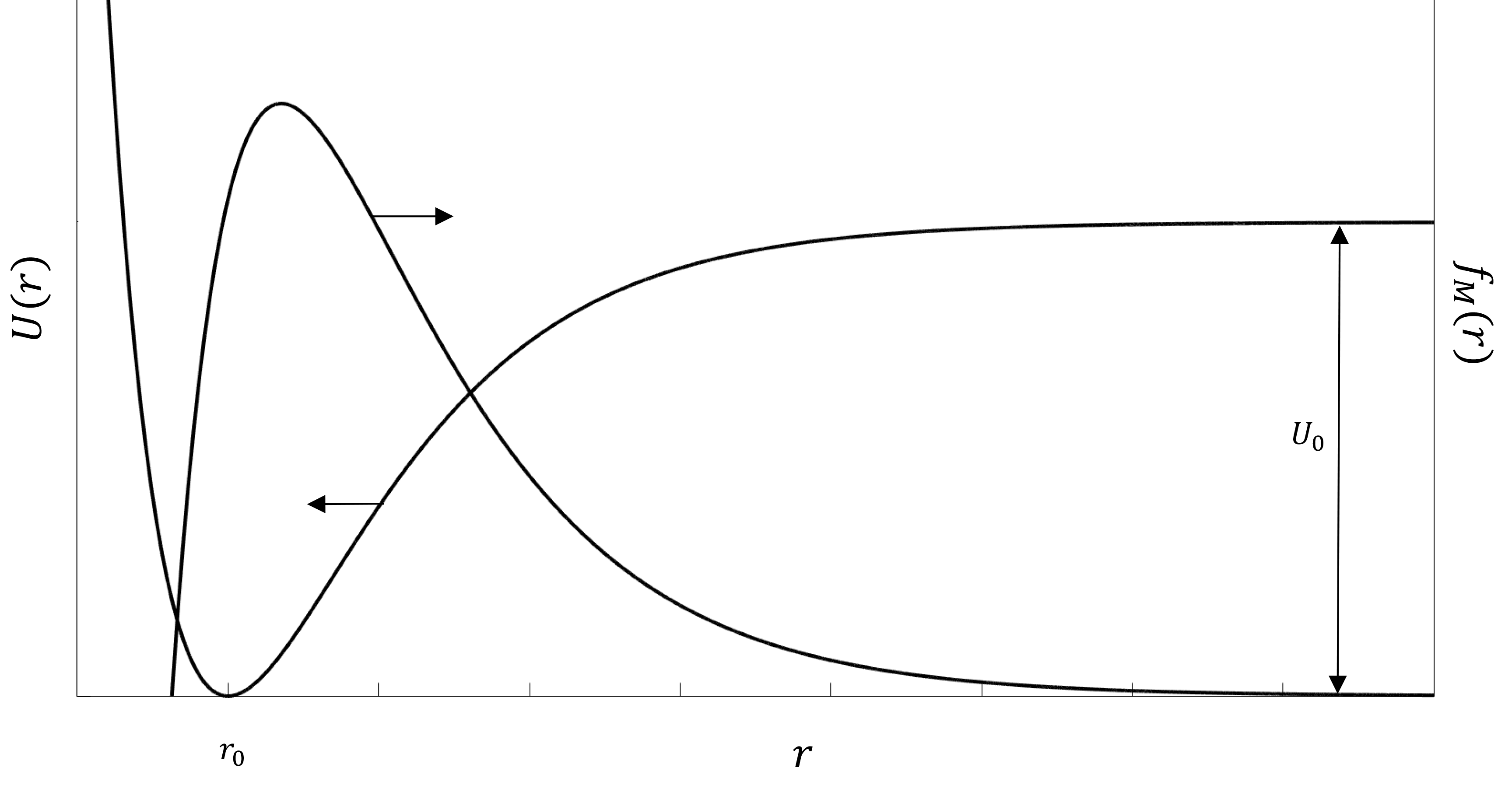}
	\caption{A Morse-type pair potential and the corresponding interatomic force.}
	\label{fig:morse}
\end{figure}

 Using (2.14) and (2.15), one can find the length of the shortest strand,  $j_{min}$, that withstands the macroscopic stretch $\lambda$ without rupturing. That is

\begin{equation} \label{eq:LeghGe}
j_{min}(\lambda)=\frac{\lambda^{2}(\theta_{0},\phi_{0})}{\xi}
\end{equation}

\noindent with

\begin{equation} 
\frac{1}{\xi}=\frac{3(3+\sqrt{4\gamma+9})}{2\gamma^{2}}+1 \qquad , \quad \gamma=\frac{\alpha a  U_{0}}{2k_{B}T}
\end{equation}

\noindent where $a$ is the characteristic length of one statistical segment. Since only elastically active strands contribute to stress production, the lower limit of the first integral in Eq. (2.11) can be replaced with $j_{min}$

\begin{equation} 
W(\lambda)=\mu\int\limits_{0}^{2\pi}\int\limits_{0}^{\pi}\int\limits_{j_{min}(\lambda)}^{\infty} \ P(j) \ w(\lambda,j) \ \sin\theta_{0} \ dj \ d\theta_{0} \  d\phi_{0}
\end{equation}

\noindent where $\mu=\frac{\sum n_{j}}{4\pi\textit{V}}$. Using the spectral decomposition theorem, the respective Cauchy stresses of the incompressible network can be derived from the strain energy density function $W(\lambda)$ as \cite{ogden1997non}

\begin{equation} 
\boldsymbol{\sigma}= \sum_{k=1}^{3} \lambda_{k} \frac{\partial W}{\partial\lambda_{k}} (\boldsymbol{n}^{(k)}\otimes \boldsymbol{n}^{(k)})  
\end{equation} \label{eq:DefStress}

\noindent where $\lambda_{k}$ and $\textbf{n}^{(k)}$ are the eigenvalues and eigenvectors of the right stretch tensor, respectively. Substitution of Eq. (2.18) into (2.19) yields

\begin{equation} \label{eq:LeghBC2}  
\boldsymbol{\sigma}= \mu\sum_{k=1}^{3} \lambda_{k}(\boldsymbol{n}^{(k)}\otimes \boldsymbol{n}^{(k)})  \  \Big(\int\limits_{0}^{2\pi}\int\limits_{0}^{\pi}  \frac{\partial}{\partial \lambda_{k}}  \int\limits_{j_{min}(\lambda)}^{\infty} \ P(j) \ w(\lambda,j)  \ \sin\theta_{0} \ dj \ d\theta_{0} \ d\phi_{0}\Big)
\end{equation}

\noindent from which all components of the Cauchy stress tensor can be evaluated (see the Appendix A).

\bigskip
 The proposed formulation can be readily generalized to include the effect of history-dependent damage in a random network subjected to a cyclic loading. From the standpoint of thermal fluctuation theory, the history-dependent damage in solids is controlled by the elementary events of bond rupture and the failure is ensued by damage accumulation in the solid. Therefore, it is assumed that the number of elastically active strands with $j$ statistical segments is a function of time, presented by $n_{j,t}$. The kinetics of irreversible bond rupture can be represented by a first-order kinetic process proposed by Eyring \cite{krausz1975deformation}

\begin{equation} 
\frac{dn_{j,t}}{dt}=-k_{r}n_{j,t} 
\end{equation}

\noindent where $k_{r}$ shows the frequency of bond rupture in elastically active strands. Using the well-known Zhurkov formula \cite{zhurkov1966kinetic}

\begin{equation} 
k_{r}=k_{r0} \ \textrm{exp}\big[f_{e}\delta/k_{B}T\big ]
\end{equation}

\noindent Here, $k_{r0}$ is a rate constant and $\delta$ is an activation length. Substitution of Eq. (2.22) into (2.21) and solving for  $n_{j,t}$ yield

\begin{equation} 
\theta_{j}(\lambda,j,t)=\frac{n_{j,t}}{n_{j,0}}=\textrm{exp}\big[\varXi(\lambda,t]\big ]
\end{equation} 

\noindent with

\begin{equation} 
\varXi(\lambda,t)=-\int\limits_{0}^{\overline{t}} \ \textrm{exp} \big[\beta(\lambda(\overline{t}))\overline{\delta} \ \big] d\overline{t}
\end{equation}

\noindent where $\overline{t}=t \ k_{r0}$ and $\overline{\delta}=\delta / a$. Here $n_{j,0}$ shows the number of elastically active strands with $j$ statistical segments before loading. Assuming

\begin{equation}
P(j)=\frac{n_{j,0}}{\sum_{j}^{} n_{j,0}}
\end{equation} 

\noindent now the time-dependent strain energy density function can be written as

\begin{equation} 
W(\lambda,t)=\mu\int\limits_{0}^{2\pi}\int\limits_{0}^{\pi}\int\limits_{j_{min}}^{\infty} \ \theta_{j}(\lambda,j,t) \ P(j)  \ w(\lambda,j) \sin\theta_{0}  \ dj \ d\theta_{0} \ d\phi_{0}
\end{equation}

\noindent with $\mu=\frac{\sum_{j}^{} n_{j,0}}{4\pi\textit{V}}$.

\section{Results}
 This section details some examples of model predictions for the elasticity and strength of polymer networks with random structure, represented by the chain length distribution (2.3). This numerical study aims to reveal how two major model parameters control the network behavior: the average length index ($\overline{j}$) and the bond strength parameter ($\xi$). Figure 2.3, shows the effect of $\overline{j}$ on the stress response of random networks subjected to quasi-static uniaxial tension and simple shear deformations. MATLAB is used to carry out the numerical calculation of integrals appearing in Eqs. (2.20) and (2.26). The assumed values for $\overline{j}$ are chosen to be comparable with the simulation results of Svaneborg  et al. \cite{svaneborg2005disorder}. Polydisperse networks with smaller $\overline{j}$ values include a larger population of short chains. Despite their slightly higher stiffness at small to moderate stretches, these networks show lower ultimate strength compared to those with larger average strand length. These conclusions can be explained considering the finite extensibility and non-Gaussian behavior of shorter strands manifested at small stretches. With increasing the applied deformation, the shorter strands gradually approach their contour length and ultimately fail under the high entropic tension. The progressive degradation of network leads to material softening and controls the ultimate strength of the network. 

\begin{figure}[H]
	\centering
	\begin{subfigure}{1\textwidth}
		\centering
		\includegraphics[width=0.9\linewidth]{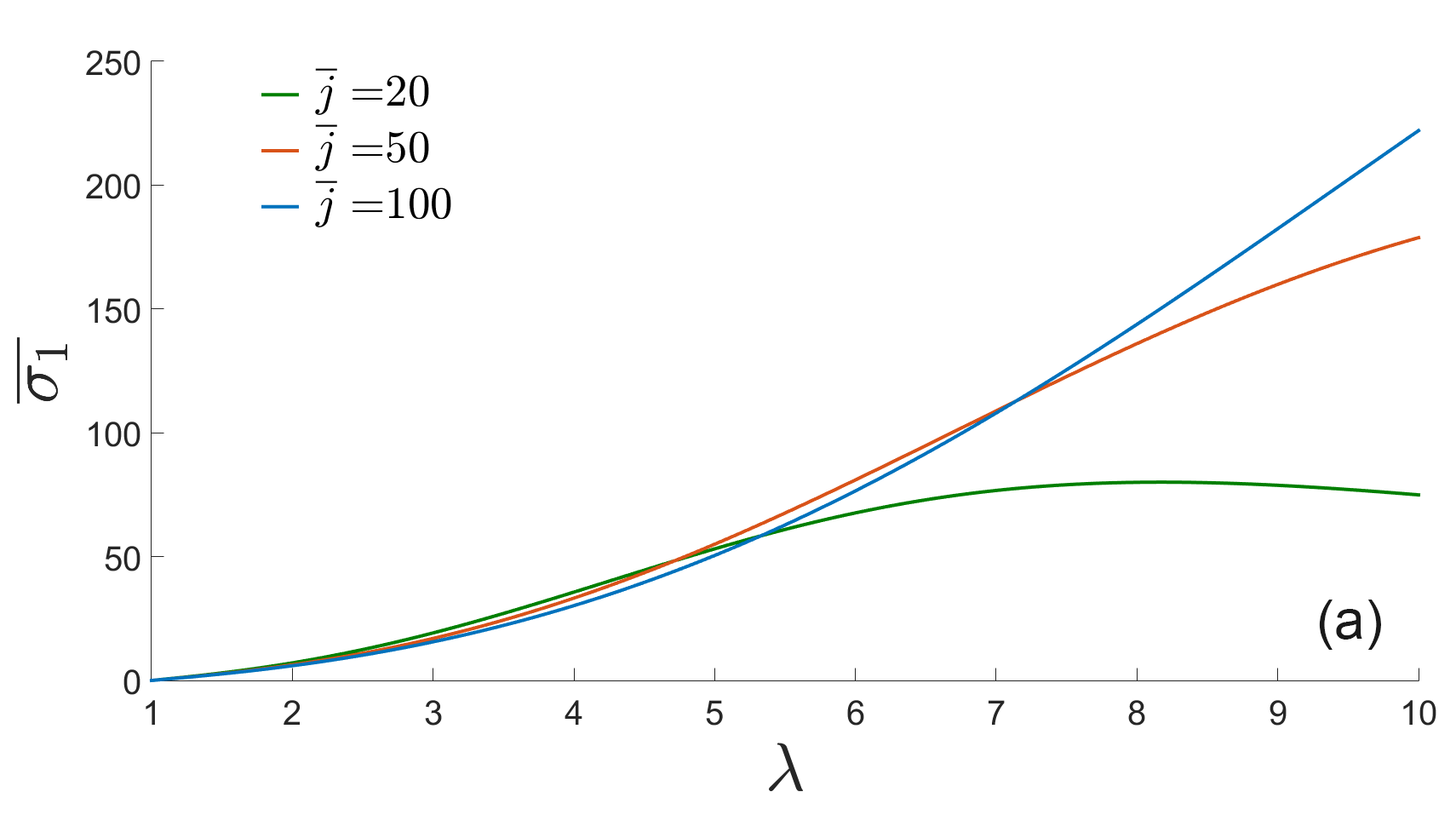}
	\end{subfigure}
	\begin{subfigure}{1\textwidth}
		\centering
		\includegraphics[width=0.9\linewidth]{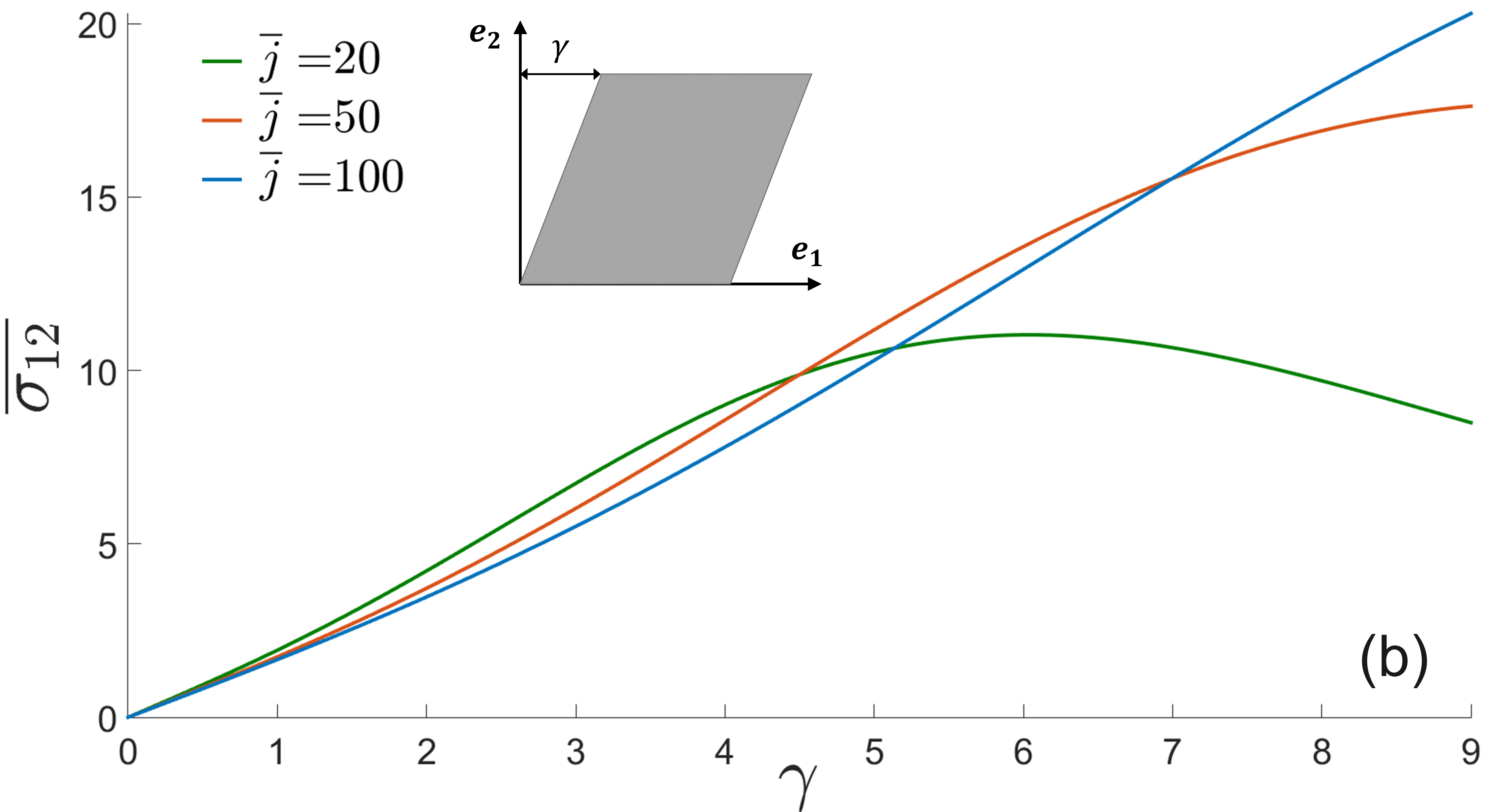}
	\end{subfigure}
	\caption{The effect of average strand length, $\overline{j}$, on the stress behavior of random networks. (a) Variation of normalized tensile stress with stretch in uniaxial tension ($\overline\sigma_{1}=\frac{\sigma_{1}-\sigma_{2}}{\mu k_{B}T}$ where $\sigma_{1}$ and $\sigma_{2}$ represent the principal stresses). (b) Variation of normalized shear stress with shear in simple shear deformation ($\overline\sigma_{12}=\frac{\sigma_{12}}{\mu k_{B}T}$). The bond strength parameter is taken to be $\xi=0.99$.}
\end{figure}

\bigskip
 The structure-properties of random networks have been the subjects of a number of molecular simulation studies \cite{grestkremer,svaneborg2005disorder,gavrilov2014computer}. The results reflect both the microstructural details and the macroscopic stress developed in the networks. Interestingly, there is a reasonable agreement between the results of  simple probability distribution (2.3) and the distribution of strand lengths in the idealized simulations \cite{grestkremer,gavrilov2014computer}. Figure 2.4 shows the stress-stretch curves in so called Mooney-Rivlin coordinates predicted by the proposed model in comparison with the simulation results of Gavrilov and Chertovich \cite{gavrilov2014computer} for random networks. They used dissipative particle dynamics to simulate the structure of randomly crosslinked polymer chains, including the effects of Langevin statistics and finite extensibility of strands. The results show an initial hardening stage due to the non-Gaussian response of the short strands, with a good agreement up to the macroscopic stretch of $\lambda\approx2.5$. The rupture of elastically active strands is a feature of the present model that is not considered in Gavrilov-Chertovich simulation. As a result, the present model predicts a drastic softening due to progressive rupture of strands whereas the predicted stress in their simulation remains practically unbounded.\par

\begin{figure}[H]
	\centering
	\includegraphics[width=0.75\linewidth]{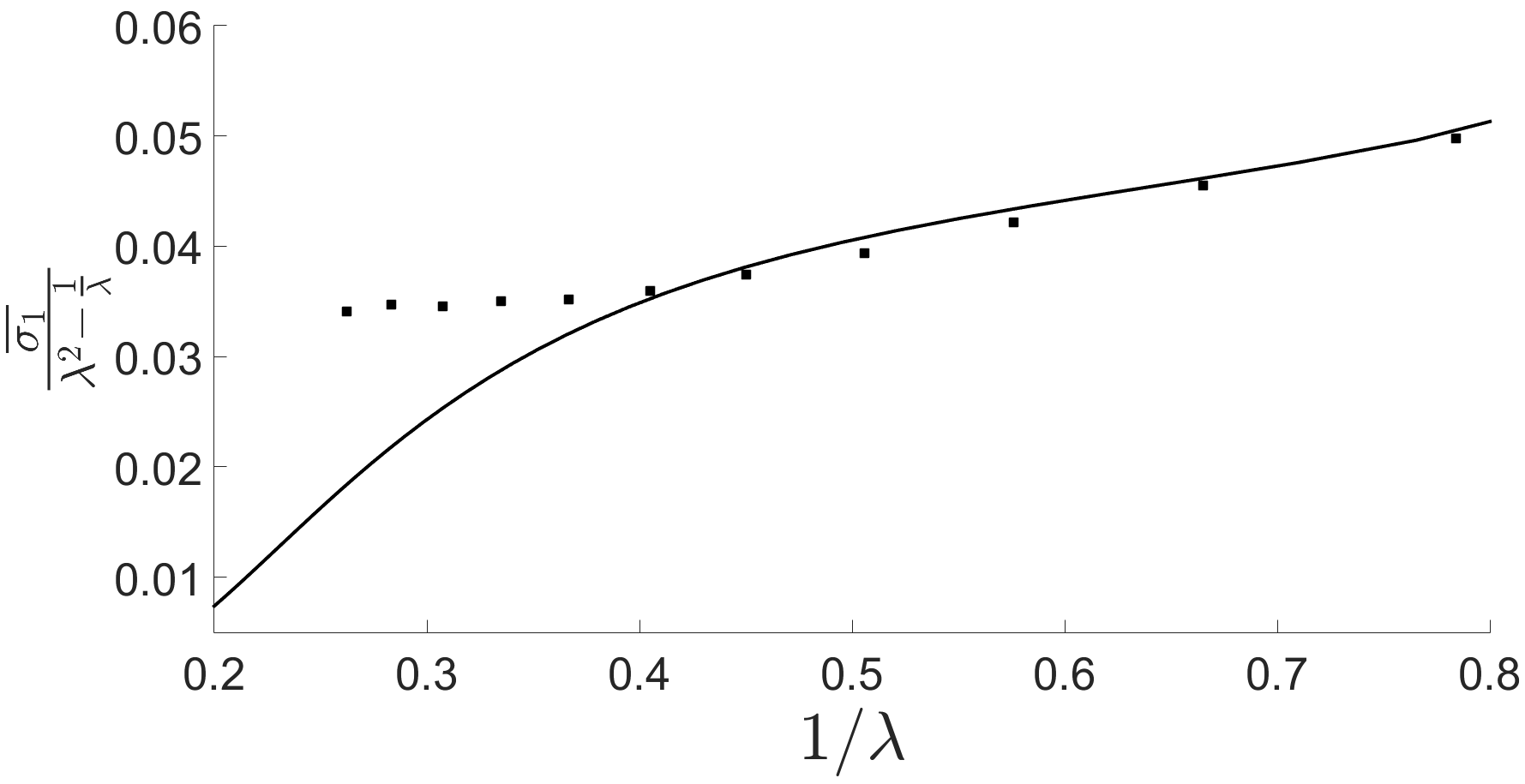}
	\caption{Comparison between the model prediction (solid line) and simulation results of Gavrilov-Chertovich \cite{gavrilov2014computer} for a random network  with $\overline{j}=10.54$ subjected to a uniaxial stress ($\overline\sigma_{1}=\frac{\sigma_{1}-\sigma_{2}}{\mu k_{B}T}$ where $\sigma_{1}$ and $\sigma_{2}$ represent the principal stresses). The bond strength parameter is taken to be $\xi=0.99$.}
	\label{fig:boat1}
\end{figure}

\bigskip
 Figure 2.5 compares the model predictions for the ultimate principal stretches (corresponding to the maximum Cauchy stress) in various plane stress loading modes  with the experimental data of Hamdi et al. \cite{hamdi2006fracture} on SBR. Seeking a generalized failure criterion at multiaxial quasi-static loadings, Hamdi et al. \cite{hamdi2006fracture} used defect-free vulcanizates and measured the elongation at break of samples subjected to uniaxial or biaxial deformations. The biaxial tests were conducted on membrane-like samples and by inflation of thin membranes in elliptical meniscuses with different aspect ratios to obtain different biaxiality ratios. The failure envelope of rubber was presented by the respective critical stretches; i.e., stretches at which the samples failed during uniaxial or biaxial loading. It appears that both $\overline{j}$ and $\xi$ are able to significantly push the failure envelope predicted by the proposed model and a wide range of experimental data could be covered by changing the value of these parameters. The strength parameter $\xi$ reflects the dissociation energy between the monomers or the crosslinking coagents. 
\begin{figure}[H]
	\centering
	\begin{subfigure}{0.7\textwidth}
		\centering
		\includegraphics[width=0.8\linewidth]{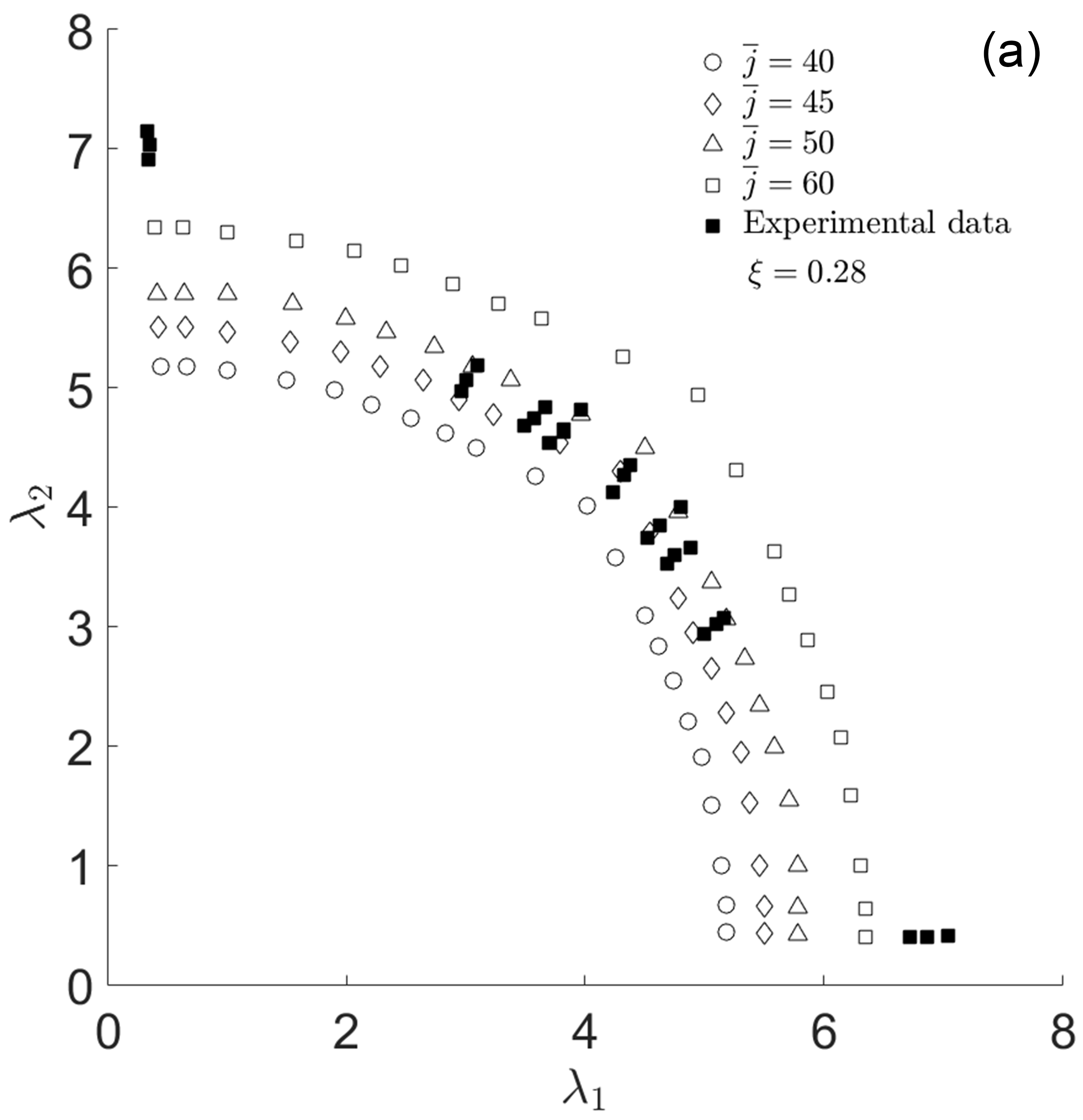}
	\end{subfigure}
	\begin{subfigure}{0.7\textwidth}
		\centering
		\includegraphics[width=0.8\linewidth]{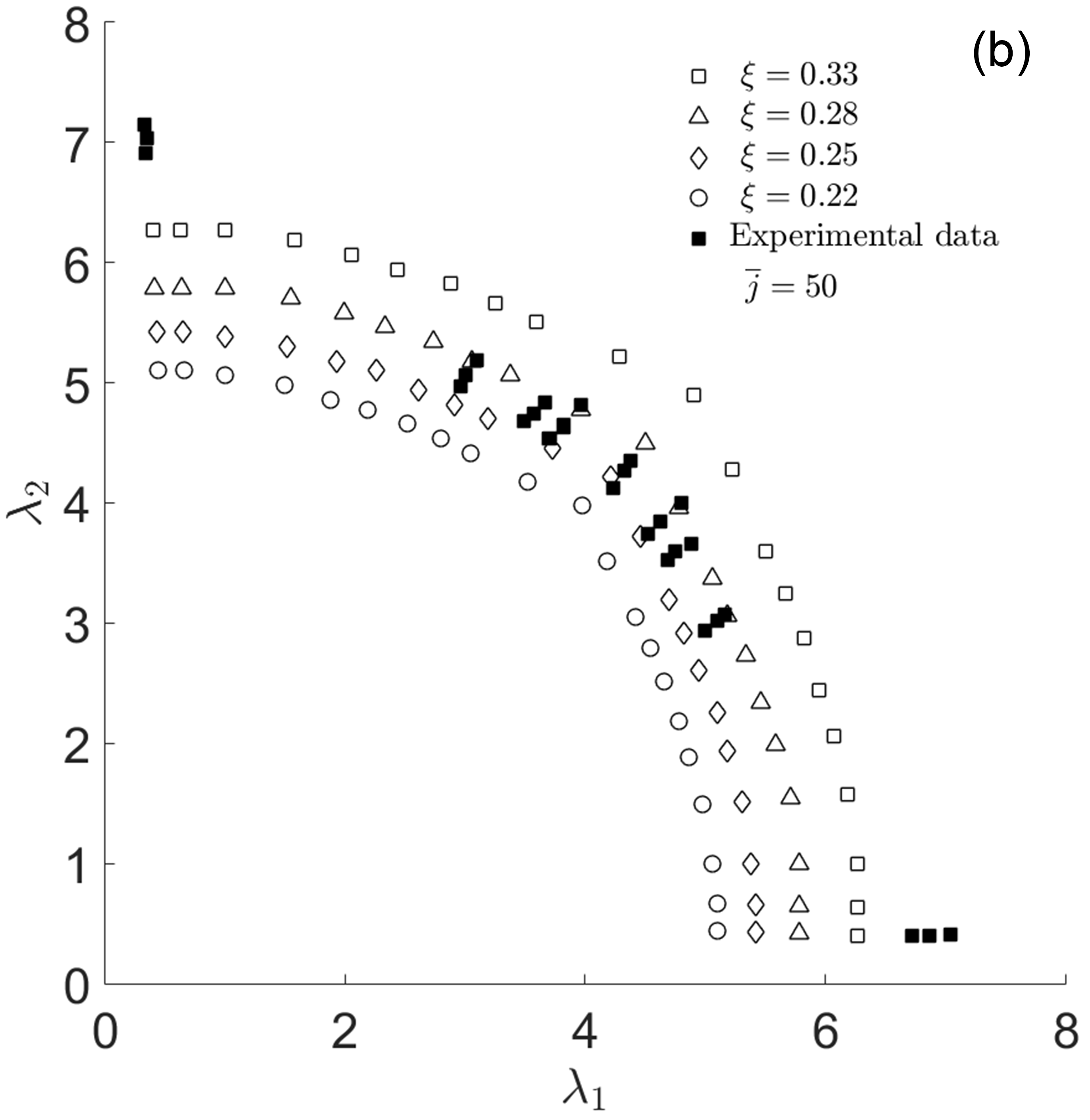}
	\end{subfigure}
	\caption{Effect of (a) average strand length, $\overline{j}$, and (b) bond strength parameter, $\xi$, on the ultimate stretches (corresponding to the maximum Cauchy stresses) as predicted by the proposed model. The results are compared with the experimental data of Hamdi et al. \cite{hamdi2006fracture} on SBR.}
\end{figure}

Figure 2.6 shows that the network strength strongly depends on $\xi$ and a small change in dissociation energy leads to a significant alteration in mechanical strength. The relative importance of networks randomness and binding energy (represented by $\overline{j}$ and $\xi$, respectively) in determination of network strength depends on the microstructural details and the nature of chemical reactions used to form the network. For example, it is well-known that carboxylated, sulfuric, and carbon-to-carbon crosslinked vulcanizates show markedly different strengths under tension \cite{bateman1963chemistry,kok1986effects}. Despite the higher dissociation energy of direct carbon-carbon bonds, however, the peroxide cures generally exhibit lower mechanical strength compared to the rubber vulcanized by accelerated sulfur \cite{gehman1969network,kok1986effects}. This strength inferiority is rooted in the significant randomness in the internal structure, introduced by peroxide reaction. Dicumyl peroxide is a vulcanizing agent that exclusively reacts with polyisoprene by abstraction of $\alpha$-methylenic hydrogen atoms. As shown by Park and Lorenz \cite{parks1963effect}, the decomposed peroxides form isoprene radicals that contribute in crosslinking with a very high efficiency. With increasing the probability of crosslinking, the population of short chains increases at the expense of strength, in accordance with the results of the presented model.\par

\begin{figure}[H]
	\centering
	\begin{subfigure}{1\textwidth}
		\centering
		\includegraphics[width=0.85\linewidth]{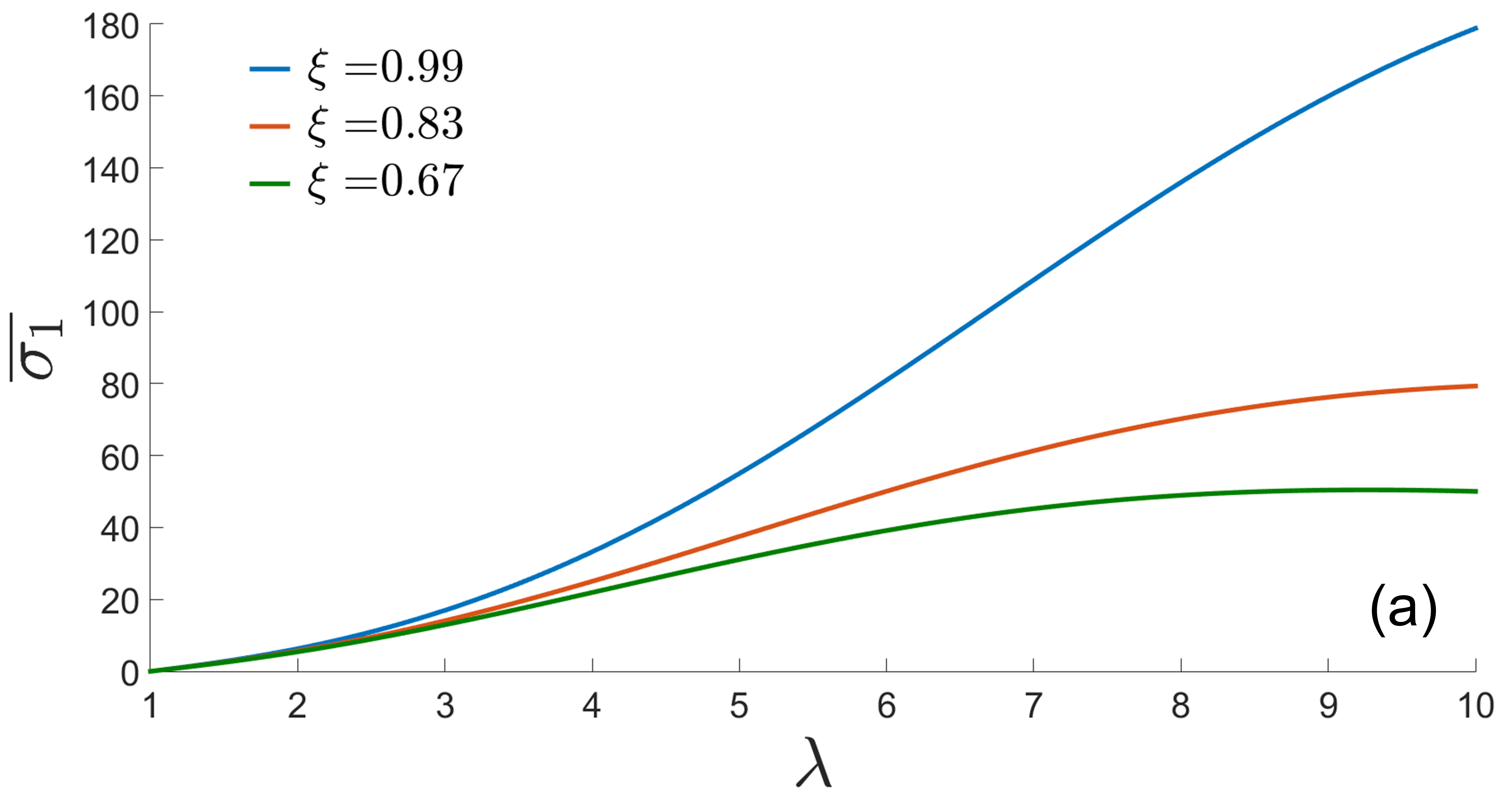}
	\end{subfigure}
	\begin{subfigure}{1\textwidth}
		\centering
		\includegraphics[width=0.85\linewidth]{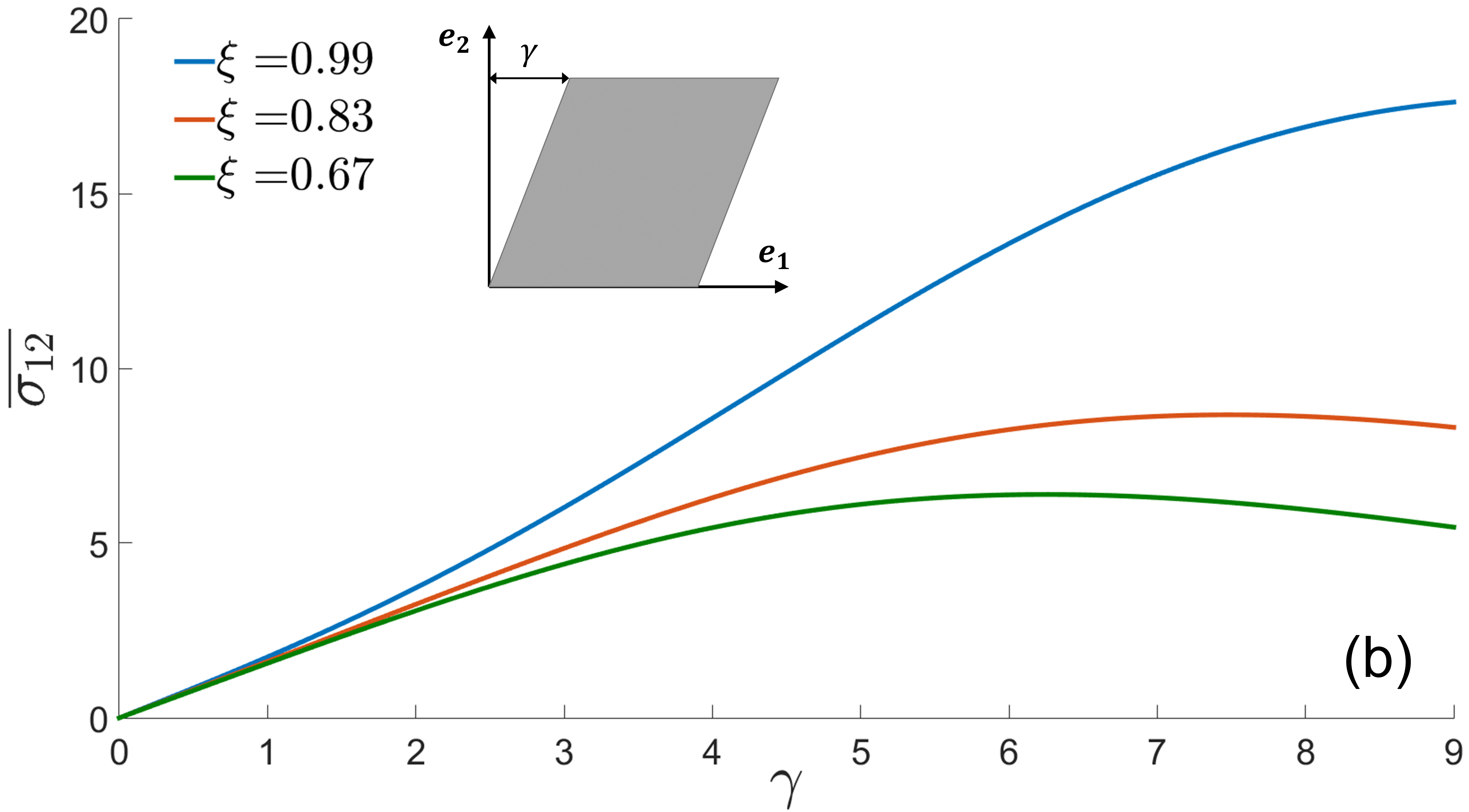}
	\end{subfigure}
	\caption{The effect of bond strength parameter, $\xi$, on the stress behavior of random networks. (a) Variation of normalized tensile stress with stretch in uniaxial tension ($\overline\sigma_{1}=\frac{\sigma_{1}-\sigma_{2}}{\mu k_{B}T}$ where $\sigma_{1}$ and $\sigma_{2}$ represent the principal stresses). (b) Variation of normalized shear stress with shear in a simple shear deformation ($\overline\sigma_{12}=\frac{\sigma_{12}}{\mu k_{B}T}$). The average strand length is taken to be $\overline{j}=20$.}
\end{figure}
\bigskip
 Finally, the model is used to predict history-dependent damage and stress-induced degradation of polydisperse networks during a cyclic loading. Figure 2.7 presents the stress-stretch behavior of two random networks with average strand length of 20 and 100. The networks are subjected to a constant amplitude periodic stretch, as shown in Figure 2.7(a). The results feature the well-known and frequently reported characteristics of rubber hysteresis \cite{ayoub2011modeling,ayoub2014visco}. The most apparent is the gradually decreasing global stiffness of the network concurrent with increasing the loading cycles. The dissipation at the first few cycles is significantly larger than the energy loss associated with the following cycles. Indeed, the hysteresis practically disappears after just a few number of deformation cycles with a constant amplitude. The energy dissipation and degradation of network mechanical properties are more pronounced in networks with smaller $\overline{j}$. These results collectively suggest that randomness in internal structure and polydispersity in strands length contribute to the fatigue behavior and could effectively limit the average lifetime of the networks.

\begin{figure}[H]
	\centering
	\begin{subfigure}{1\textwidth}
		\centering
		\includegraphics[width=0.85\linewidth]{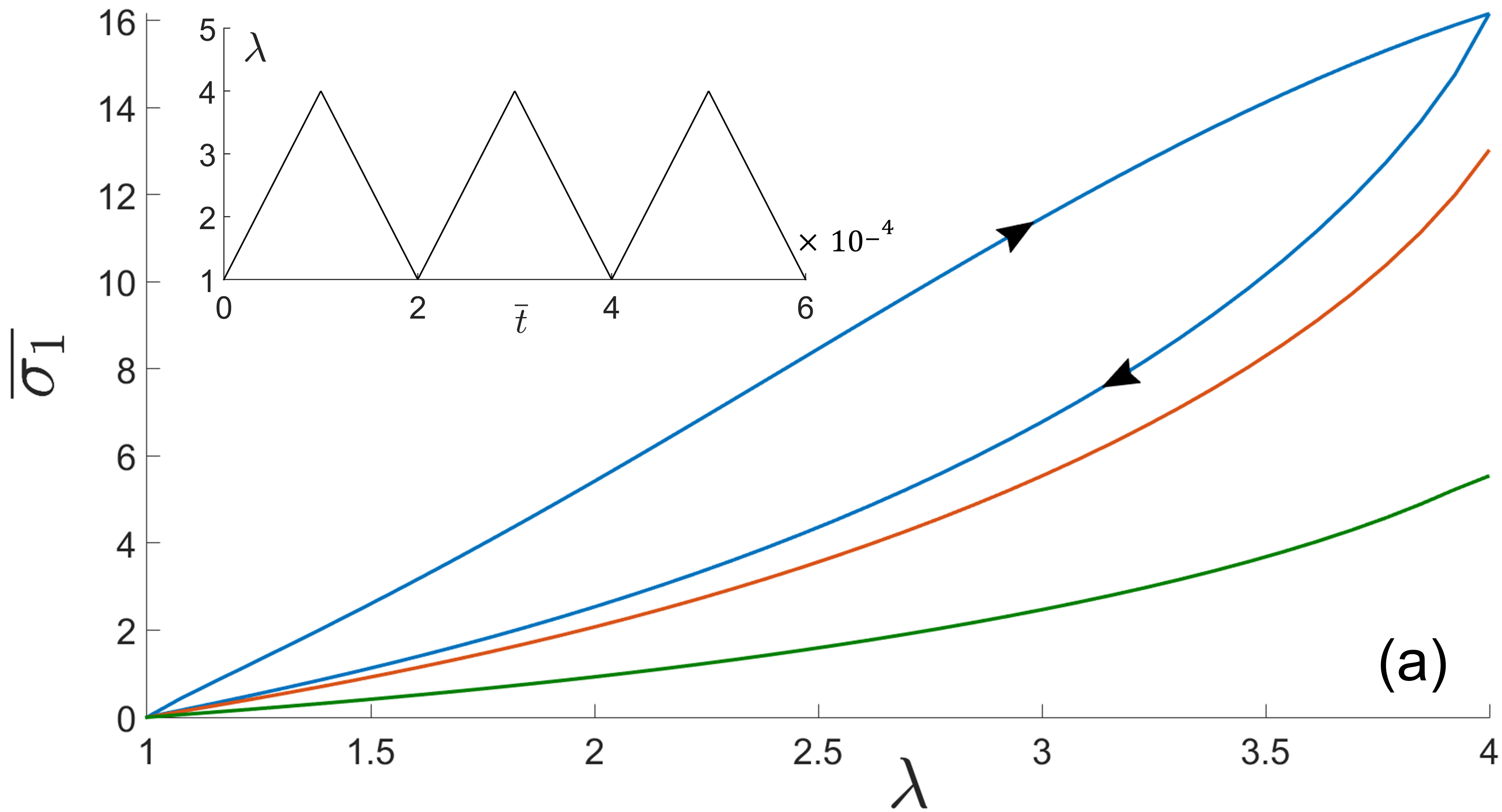}
	\end{subfigure}
	\begin{subfigure}{1\textwidth}
		\centering
		\includegraphics[width=0.85\linewidth]{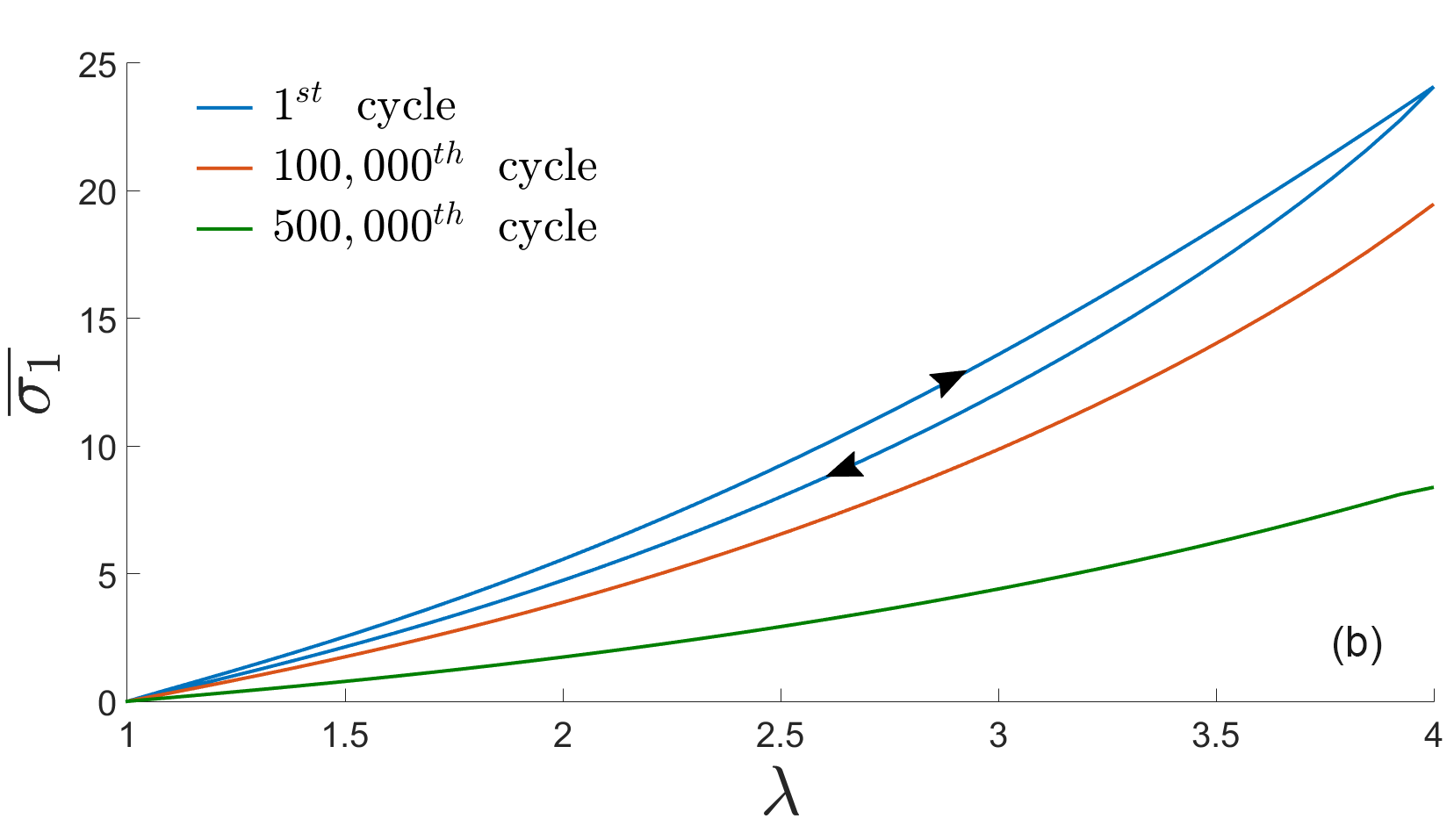}
	\end{subfigure}
	\caption{Variation of normalized stress with cyclic stretch in random networks with (a) $\overline{j}=20$ (b) and $\overline{j}=100$ ($\overline\sigma_{1}=\frac{\sigma_{1}-\sigma_{2}}{\mu k_{B}T}$ where $\sigma_{1}$ and $\sigma_{2}$ represent the principal stresses). The networks are subjected to slow cyclic axial stratching, as shown by inset. A full cycle of loading-unloading lasts 50 $\textrm{s}$. Other model parameters are $\xi=0.99$, $k_{r0}=2\times10^{-6} \ \textrm{s}^{-1}$ \cite{lavoie2}, and $\delta \approx a$ \cite{yarin1998model}.}
\end{figure}

\section{Concluding Remarks}

 In an attempt to correlate the ultimate macroscopic mechanical properties of polymeric networks to their internal structure, a theory of rubber elasticity is formulated in which the network microstructure is random and the strands are polydisperse in length. On the basis of a simple statistical analysis, new expressions for the strain energy density function and the Cauchy stress tensor are obtained that take into account the strand length distribution and predict its effect on the bulk damage in the network. Strands with different lengths respond differently to the applied macroscopic deformation. Short chains quickly experience the Langevin effect and break under a relatively small stretch. The progressive failure of the shorter strands continues and eventually determines the ultimate strength of the network. Direct mechanical measurements are insufficient to exclusively provide any information on the strand length distribution and the randomness in internal structure of polymer networks. Thus, the value of presented model is that it can be used to test the validity of assumptions made about the network statistics at the microscale by comparing the model predictions with the relevant experimental data.

\bigskip
 Certain remarks must be made with regard to the validity and capability of the proposed approach. First, the validity of simple statistical model presented by Eq. (2.3) depends on a major assumption that all statistical segments have an equal chance to contribute to crosslinking reactions. While this assumption may be acceptable for peroxide cures, it does not do justice to the complicated structure of sulfur vulcanizates. Vulcanization of natural rubber with accelerated sulfur is essentially an autocatalytic reaction \cite{gehman1969network,alfrey1963kinetics,arends1963general}. Sulfur facilitates the local reactions adjacent to a crosslink and leads to increased functionality and further enhancement of the network strength. Second, the assumption that all strands in a polydisperse network follow an affine deformation is not backed by a rigorous justification. Very short strands, say with just a few statistical segments in length, hardly act as elastically active chains \cite{tobolsky,bueche} and thus their failure under small deformation is unlikely. This becomes particularly important when considering the distribution function (2.3) in which network strands of short size could be in abundance. Third, while the energetic Morse potential is used to predict the bond rupture, the contribution of enthalpic interactions in the free energy (Eq. (2.6)) is disregarded. It is known that the consideration of enthalpic contributions removes the singularity caused by the Langevin effect at large deformations. Inclusion of enthalpic contributions in calculation of free energy is expected to provide a more realistic estimation of the network strength.

\chapter[Revisiting the Stress-Induced Damage in Filled Elastomers: Effect of Polydispersity]{Revisiting the Stress-Induced Damage in Filled Elastomers: Effect of Polydispersity\footnote{This manuscript version is made available under the CC-BY-NC-ND 4.0 license,  http://creativecommons.org/licenses/by-nc-nd/4.0/}}

\section{Abstract}
 A priori assumption in micromechanical analysis of polymeric networks is that the constitutive polymer strands are of equal length. Monodisperse distribution of strands, however, is merely a simplifying assumption. In this paper, we relax this assumption and consider a vulcanized network with a broad distribution of strand length. In the light of this model, we predict the damage initiation and stress-stretch dependency in filled polymer networks with random internal structures. The degradation in network mechanical behavior is assumed to be controlled by the adhesive failure of the strands adsorbed onto the filler surface. We show that the finite extensibility of the short adsorbed strands is a key determinant of mechanical strength.

\section{Introduction}
 Small filler particles like carbon black, silica, and clay are often compounded with rubbers to improve their mechanical properties, including stiffness, abrasion resistance, tenacity, and durability \cite{hamed2000reinforcement,medalia1987effect,waddell1996use}. The fillers are also the major contributors to the damage nucleation and underlie the stress and strain softening mechanisms in filled elastomers. History dependence, or the \textquotedblleft Mullins effect \textquotedblright, is a particular feature of the mechanical response of filled elastomers in which the material shows hysteresis during quasi-static loading and softens with the history of loading. Another prominent example is the amplitude dependence of viscoelastic moduli of filled rubbers. Elastomers generally show linear viscoelastic properties in strain amplitudes up to 20 \cite{chazeau2000modulus}. The storage modulus of elastomers loaded with solid particles, however, shows a large drop with increasing strain. This strain-softening phenomenon is often referred to as the “Payne effect” \cite{payne1962dynamic1,payne1962dynamic2,payne1965reinforcement}.

\bigskip
 Great interest has been kindled in understanding the origin of damage initiation in mechanical behavior of filled elastomers due to its great practical importance in tire industry \cite{clark1978rolling,leblanc2009filled}. Over the past decades, different micro-mechanical and phenomenological mechanisms are proposed to explain the mechanism of damage nucleation and growth in filled elastomers \cite{leblanc2002rubber}. As initially noted by Payne and Kraus \cite{kraus1984mechanical,payne1972effect}, degradation of mechanical properties may arise from disruption of the agglomerated particles. Strong inter-particle interaction among surface active particles, like silica, leads to formation of disorderly grown aggregates with fractal structure, ranging from 10 to 100 nm in size \cite{kluppel2003role}. Increasing the filler content beyond the percolation threshold creates filler networks at larger scales within the matrix. Large strain perturbations deform and eventually disrupt the  aggregates and introduce strong nonlinearity in the mechanical behavior of rubber composites \cite{heinrich2002recent}. 

\bigskip
 In a different approach, the alteration of networks mechanical properties is ascribed to the nature of rubber-filler interactions. Polymer molecules generally show affinity for the surface of active particles. This is mediated either by chemical and strong physical bonds \cite{stockelhuber2011impact,leopoldes2004influence,ramier2007payne} or by disorder-induced localization of polymer onto the rough surface of particles \cite{vilgis1994disorder,vilgis2005time}. The structure of adsorbed polymer layer changes with applied deformation. Bueche \cite{bueche} argued that deformation of a filled rubber will break the highly stretched chains bridging the two adjacent fillers or tear them loose from the filler surface. Maier and G\"{o}ritz \cite{maier1996molecular} proposed a different mechanism, in which the affinity between fillers and polymer chains favors the establishment of stable and unstable bonds on the filler surface. The adhesion sites between the polymer chains and fillers are regarded as temporary and supplemental “crosslink points” contributing to the entropic elasticity. Unstable bonds have less drag-resistance and break under the elevated interfacial stresses \cite{raos2013pulling}. Large deformations promote the polymer disentanglement from filer surface causing the overall stiffness to drop. More recently, Jiang \cite{jiang2014effect} showed this Langmuir-type desorption of polymer chains could be due to the non-uniform stretching of chains in the matrix.

\bigskip
 Most of the foregoing phenomenological descriptions have been integrated into micromechanical material models for filled rubbers (see Govindjee and Simo \cite{govindjee1991micro} and Heinrich and Kl\"{u}ppel \cite{heinrich2002recent} and references therein). The goal of present contribution is to add a new dimension to the constitutive modeling of bulk damage in filled vulcanizates by accounting for the random internal structure of rubber and the polydispersity of strands. Monodispersity of constitutive strands has been taken as a priori assumption in classical theory of rubber elasticity. The vulcanization of rubbers, however, is an inherently random process which likely results in formation of polydisperse networks. The importance of strand length distribution in mechanical behavior of vulcanizates was first recognized in the pioneering works of Bueche \cite{bueche}, Watson \cite{watson1953chain,watson1954chain}, and Gehman \cite{gehmanmolecular}. Models for mechanical behavior of polydisperse networks that take into account the statistical information of the strand length distribution are developed only recently \cite{itskovrubber}. In this paper, we will examine the effect of polydispersity on elasticity and mechanical strength of filled vulcanizates with polydisperse structures. We develop a micromechanical model for initiation of bulk damage and show how the irregularity in network structure markedly affects the strength of filled rubbers.

\section{Model}

 Bueche \cite{bueche} and Watson \cite{watson1953chain,watson1954chain} first proposed a simple distribution function for strand length in a random polymer network. Accordingly, if ${n_{j}}$ represents the number of strands with ${j}$ statistical segments, the probability distribution of having a strand with ${j}$ statistical segments, $P(j)$, can be expressed as

\begin{equation} 
	P(j)=\frac{1}{\overline{j}}e^{-j/\overline{j}}
\end{equation}

\noindent where $\overline{j}=\frac{1}{p}$ and ${p}$ shows the probability of a segment to be crosslinked. In deriving Eq. (3.1), it was assumed that the placement of crosslinks on the main chain is completely random, all segments have an equal chance ${p}$ to participate in vulcanization, and the segmental bindings during crosslinking are statistically independent events. $\overline{j}$ is a decay length (Figure 3.1) and represent the average strand length of the random network. 

\begin{figure}[H]
	\centering
	\includegraphics[width=0.85\linewidth]{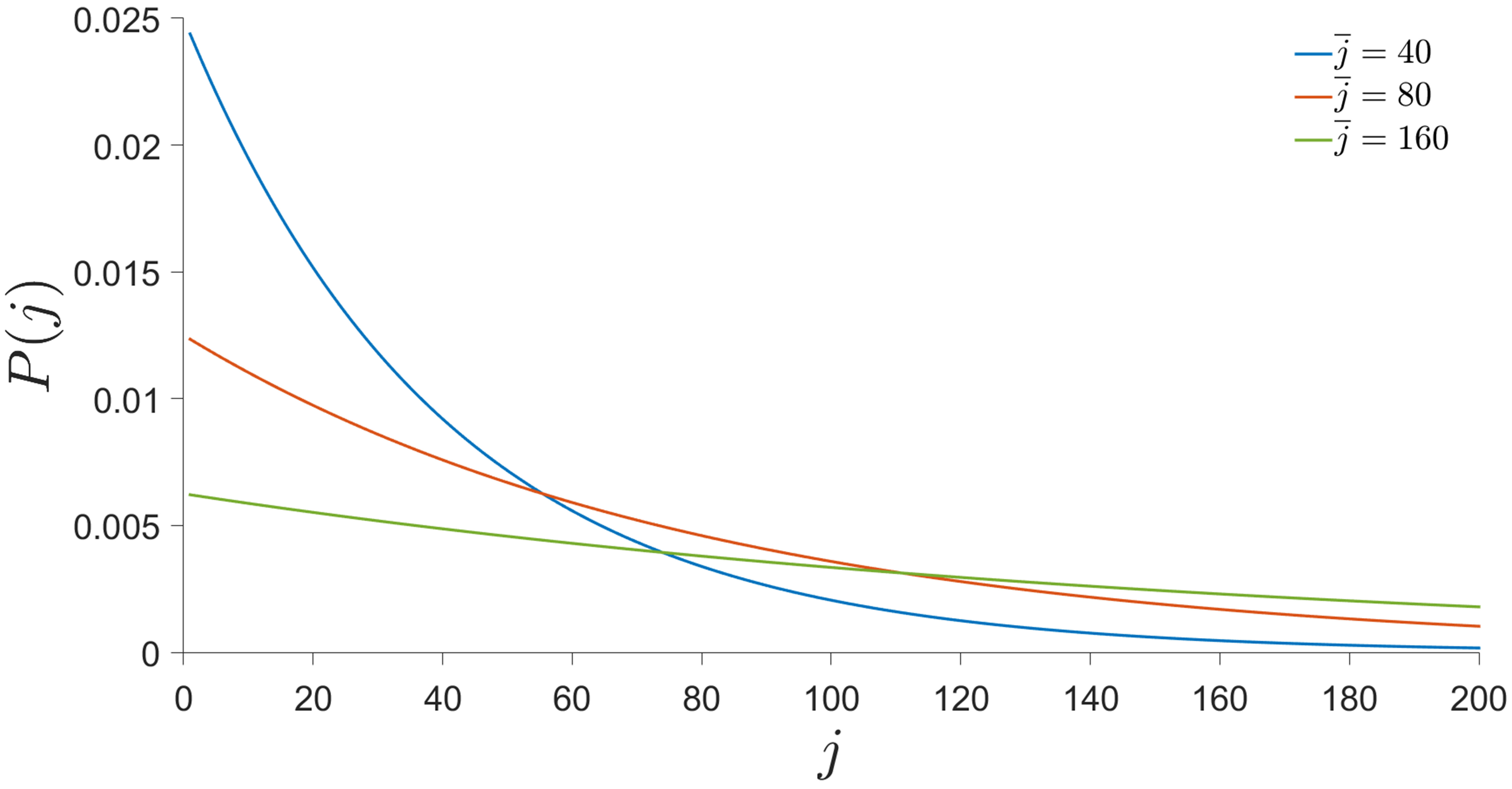}
	\caption{Probability distribution of strands.}
	\label{fig:boat1}
\end{figure}

 Consider a random dispersion of rigid fillers, with volume fraction $\nu_{f}$ in an elastomeric matrix, with a strand length distribution that follows Eq. (3.1). The polymer strands close to the surface of particles may reversibly interact with the affine particle surface and establish labile bonds (Figure 3.2). This affinity leads to formation of a transition zone, with volume fraction $\nu_{f}$ around each particle in which segments of strands may adsorb (desorb) to (from) the particles. Upon adsorption, each strand with length ${j}$ is divided into two sub-chains with shorter length. For simplicity, we assume that the sub-chains have similar length and include ${j/2}$ statistical segments. 

\begin{figure}[H]
	\centering
	\includegraphics[width=0.85\linewidth]{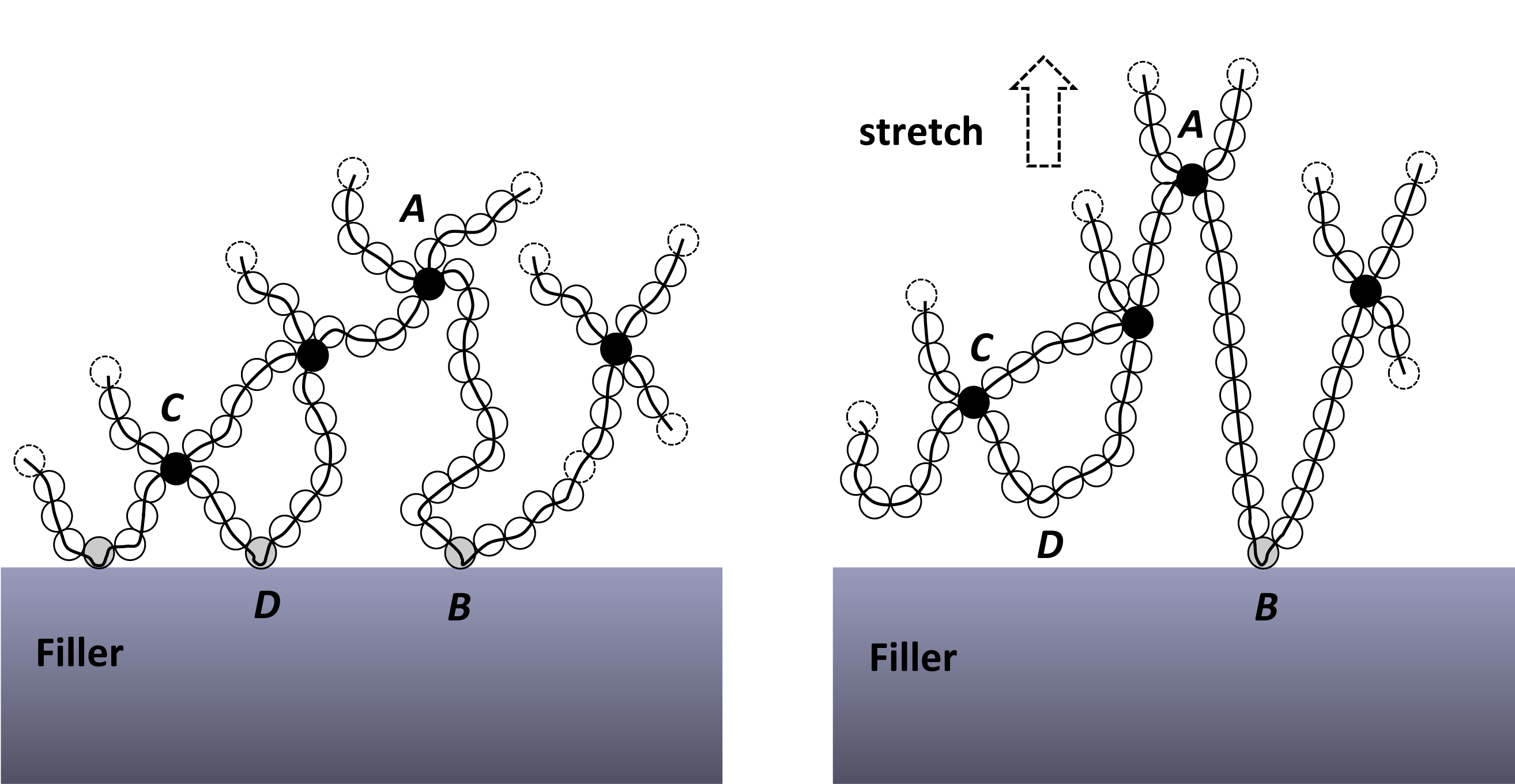}
	\caption{A polydisperse network close to the surface of a filler. The circles schematically show the statistical segments forming the polymer chains. (Left) before application of macroscopic deformation, polymer strands may attach to the filler surface and form physical bonds at multiple points (B and D). AB and CD show two adsorbed strands with different lengths. (Right) as the system deforms, shorter strands (e.g., CD) desorb from the filler surface due to relatively larger entropic forces developed in them.}
	\label{fig:boat1}
\end{figure}

\bigskip
 Let $n_{j}^{(i)}(t)$ show the number density of the strands at time $t$ where the superscripts $j=a, f$, and $b$ refer to the adsorbed strands in the transition zone, free strands in the transition zone, and free strands in the bulk, respectively. The competitive adsorption/desorption of strands in the adhesion zone can be represented by the following kinetic equations

\begin{subequations}
	\label{eq:gen-dens-rel}
	\begin{align}
	&\frac{d n_{j}^{(a)}(t)}{d t}=-k_{r}n_{j}^{(a)}(t)+k_{f}n_{j,t}^{(f)}\\
	&\frac{d n_{j,t}^{(f)}}{d t}=-k_{f}n_{j,t}^{(f)}+k_{r}n_{j}^{(a)}(t)\\
	&\frac{1}{2} n_{j}^{(a)}(t)+n_{j}^{(f)}(t)=n_{j}^{(b)}(t)
	\end{align}
\end{subequations}

\noindent where $k_{f}$ and $k_{r}$ stand for the forward and reverse rates of strand adhesion to the filler surface. Under a static or quasi-static loading condition, the concentrations of strands maintain their steady state values at

\begin{subequations}
	\label{eq:gen-quasi-rel}
	\begin{align}
	&n_{j}^{(a)}=\frac{\kappa n_{j}^{(b)}}{1+\frac{\kappa}{2}} \\
	&n_{j}^{(f)}=\frac{n_{j}^{(b)}}{1+\frac{\kappa}{2} } 
	\end{align}
\end{subequations}

\noindent where $\kappa=\frac{k_{f}}{k_{r}}$.

\bigskip
 The composite system described above is subjected to a homogeneous deformation gradient $\textbf{F}$. The end-to-end vector of each representative strand at the reference configuration is shown by $\textbf{R}^{(i)}_{0}$. After deformation, the strand finds a new configuration, shown by the end-to-end vector $\textbf{R}^{(i)}$. The strands are assumed to be flexible chains formed by $j$ freely joined statistical segments. Taking into account the Langevin effect, the stored energy of each strand, stretched by $\lambda_{e}$, is \cite{arruda1993three}

\begin{equation} 
w^{(i)}(\lambda,j)=jk_{B}T\bigg (\frac{\lambda_{e}\beta}{\sqrt{j}}+\ln\frac{\beta}{\sinh\beta} \bigg)+w^{(j)}_{0}
\end{equation}

\noindent where $k_{B}T$ is the thermal energy, $w_{0}$ represents the deformation-independent part of the free energy, and 

\begin{equation} 
\beta=\pounds^{-1}\bigg (\frac{\lambda_{e}}{\sqrt{j}}\bigg ) \quad
\end{equation}

\noindent with $\pounds^{-1}$ being the inverse Langevin function. Here, $\lambda_{e}$ refers to the magnitude of \textquotedblleft effective \textquotedblright stretch experienced by each strand. Following Boyce and Qi \cite{qi2004constitutive}, we assume that inclusions amplify the matrix stretch in the matrix and the effective stretch experienced by strands is somewhat larger than the far-field stretch $\lambda$. Boyce and Qi approximated this amplification using

\begin{equation}\label{eq:AmpStr}
\lambda_{e} =1+X(\lambda-1)
\end{equation}

\noindent where

\begin{equation}\label{eq:AmpStr}
X=1+2.5\nu_{f}+14.1\nu_{f}^{2}
\end{equation}

\noindent is the hydrodynamic correction factor.

 The endpoint of the representative strand at the reference configuration is located on a sphere with radius of $R^{(i)}_{0}$, with spherical coordinates ${(R^{(i)}_{0},\theta_{0},\phi_{0})}$. If the medium is undeformed in the reference configuration, we may assume that the strands are randomly oriented in space before deformation. Thus, the number of strands with $j$ statistical segments whose endpoints falls in ${(R^{(i)}_{0},\theta_{0}+d\theta_{0},\phi_{0}+\phi_{0})}$ is given by 

\begin{equation} 
dn^{(i)}_{j}=\frac{1}{4\pi}n^{(i)}_{j} sin\theta_{0} \ d\theta_{0} \ d\phi_{0}
\end{equation}

\noindent Hence, the total free energy of strands with $j$ statistical segments is

\begin{equation} 
W^{(i)}_{j}(\lambda)= n^{(i)}_{j}\int\limits_{0}^{2\pi}\int\limits_{0}^{\pi} w^{(i)} \big (\lambda_{e},j \big ) \  \sin\theta_{0} \ d\theta_{0} \ d\phi_{0}
\end{equation}

\noindent The stretch along an arbitrary direction can be expressed in the reference configuration and in terms of the macroscopic principal stretches, $\lambda_{i}$, as

\begin{equation}
\lambda^{2} (\theta_{0},\phi_{0})=(\lambda_{1}  \sin\theta_{0}  \cos\phi_{0})^{2}+(\lambda_{2}  \sin\theta_{0}   \sin\phi_{0})^{2}+(\lambda_{3}  \cos\theta_{0})^{2}
\end{equation}

\noindent The total free energy of the entire population of chains, $W^{(i)}\big (\lambda_{e})$ is the summation of contribution of individual chains. That is, $W^{(i)}(\lambda_{e})=\sum W^{(i)}_{j}(\lambda_{e})$. Considering a continuous distribution for polydisperse chains, as shown by Eq. (3.1), the summation can be replaced by an integral. For the bulk strands, it leads to

\begin{equation}
W^{(b)}(\lambda)=\mu\int\limits_{0}^{2\pi}\int\limits_{0}^{\pi}\int\limits_{1}^{\infty} \ P^{(b)}(j) \ w \big (\lambda,j \big )  \sin\theta_{0} \ dj \ d\theta_{0} \ d\phi_{0}
\end{equation}

\noindent where 

\begin{equation} 
P^{(b)}(j)=\frac{1}{\overline{j}^{(b)}}e^{-j/\overline{j}^{(b)}}
\end{equation}

\noindent and $\mu=\frac{\sum n_{j}}{4\pi}$. Similar equations can be derived for the free and adsorbed chains in the transition zone

\begin{subequations}
	\label{eq:gen-quasi-rel}
	\begin{align}
		&W^{(f)}(\lambda)=\mu\int\limits_{0}^{2\pi}\int\limits_{0}^{\pi}\int\limits_{1}^{\infty} \ P^{(f)}(j) \ w \big (\lambda,j \big )  \sin\theta_{0} \ dj \ d\theta_{0} \ d\phi_{0}\\
		&W^{(a)}(\lambda)=\mu\int\limits_{0}^{2\pi}\int\limits_{0}^{\pi}\int\limits_{1}^{\infty} \ P^{(a)}(j) \ w \big (\lambda,j \big )  \sin\theta_{0} \ dj \ d\theta_{0} \ d\phi_{0}
	\end{align}
\end{subequations}

\noindent $P^{(f)}(j)$ and $P^{(a)}(j)$ in Eqs. 3.13(a,b) are defined by the following ratios 

\begin{subequations}
	\label{eq:gen-quasi-rel}
	\begin{align}
		P^{(f)}(j)=\frac{n^{(f)}_{j}}{\sum_{j}^{}n_{j}}\\
		P^{(a)}(j)=\frac{n^{(a)}_{j}}{\sum_{j}^{}n_{j}}
	\end{align}
\end{subequations}

 Note that these two quantities do not represent the probability distribution of the free and adsorbed chains. They are simply the ratio of population free and adsorbed strands with $j$ segments within the transition zone to the number of total strands.

\bigskip
 Using Eq. (4), the average entropic force developed in a strand with $j$ segments can be obtained as

\begin{equation} 
f_{j}(\lambda)= \frac{k_{B}T\beta}{l}
\end{equation}

\noindent where $l$ is a characteristic length of a statistical segment. Eq. (3.15) accounts for the finite extensibility of strands and thus diverges as the stretch approaches the ultimate value of $\lambda_{lock}=\sqrt{j}$ \cite{arruda1993three}. Since the network is assumed to move affinely, shorter strands experience larger entropic tension even under a small macroscopic stretch. These large entropic forces lead to microscopic degradation of the filled network. The entropic tension of strands shortens the lifetime of the physical bonds. If tension is strong enough, free or bulk strands break at backbone or cleave at a crosslink and become elastically inactive. To take the deformation induced network alteration into account, we follow the method proposed by Itskov and Knyazeva \cite{itskovrubber} and replace the lower limit of the first integral in Eq. (3.11) and (3.13) with the shortest strand that can survive the macroscopic stretch. The dissociation energy of labile or covalent bonds is modeled using Morse pair-potentials \cite{crist1984polymer,dal2009micro}

\begin{equation} \label{eq:LeghGe}
U^{(i)}(r)=U^{(i)}_{0}\Big(1-\textrm{exp}\big[-\alpha^{(i)}(r^{(i)}-r^{(i)}_{0})]\Big)^2
\end{equation}

\noindent where $U^{(i)}_{0}$ is the dissociation energy and $\alpha^{(i)}$ is a constant that determines bonds elasticity. $r^{(i)}$ and $r^{(i)}_{0}$ show the deformed and equilibrium length of a bond, respectively. The dissociation or rupture of strands occurs when the maximum entropic force developed in a strand reaches the critical value of $\frac{\alpha U^{(i)}_{0}}{2}$. Using Eq. (3.15), we can find $j^{(i)}_{min}$, the number of statistical segments of the shortest elastically active strand that survives the effective macroscopic stretch $\lambda_{e}$ as

\begin{equation} \label{eq:LeghGe}
j^{(i)}_{min}(\lambda)=\frac{\lambda^{2}_{e}(\theta_{0},\phi_{0})}{\xi^{(i)}}
\end{equation}

where 

\begin{equation} 
\frac{1}{\xi^{(i)}}=\frac{3(3+\sqrt{4\gamma^{(i)}+9})}{2(\gamma^{(i)})^{2}}+1 \qquad , \quad \gamma^{(i)}=\frac{\alpha^{(i)} U^{(i)}_{0} l }{2k_{B}T}
\end{equation}

\noindent where $l$ is the characteristic length of one statistical segment. Note that $\xi^{(b)}=\xi^{(f)}$. Replacing the lower limit of the first integral in Eq. (3.11) and (3.13) with $j^{(i)}_{min}$, we obtain

\begin{equation} 
W^{(i)}(\lambda)=\mu\int\limits_{0}^{2\pi}\int\limits_{0}^{\pi}\int\limits_{j^{(i)}_{min}(\lambda)}^{\infty} \ P^{(i)}(j) \ w(\lambda_{e},j) \ \sin\theta_{0} \ dj \ d\theta_{0} \  d\phi_{0}
\end{equation}

\noindent The detachment of adsorbed strands changes the balance of their steady state density in the transition zone. We consider this effect using a Zhurkov-type unbinding rate as \cite{zhurkov1966kinetic}

\begin{equation} 
k_{r}=k^{0}_{r}\textrm{exp} \ \big[f_{j}\delta\diagup k_{B}T\big]
\end{equation}

\noindent where $k^{0}_{r}$ is a constant and $\delta$ is an activation length.

\bigskip
 The average of total strain energy stored in the matrix can be represented as 

\begin{equation} 
	W=\upsilon_{b} W^{(b)}+\upsilon_{t}(W^{(f)}+W^{(a)})
\end{equation}

The stresses produced in the incompressible network can be derived from the strain energy density, using the spectral decomposition theorem

\begin{equation} 
	\boldsymbol{\sigma}= \lambda_{k} \frac{\partial W}{\partial\lambda_{k}} (\boldsymbol{n}^{(k)}\otimes \boldsymbol{n}^{(k)})  
\end{equation} \label{eq:DefStress}

\noindent where $\lambda_{k}$ and $\textbf{n}^{(k)}$ are the eigenvalues and eigenvectors of deformation tensor, respectively.

\section{Results}
 In this section, we present some numerical examples illustrating the effect of polydispersity on the overall static behavior of a filled network. We explicitly focus on the effect of average length index ($\overline{j}$), representing the randomness in network structure, and the strength parameter ($\xi^{(i)}$). To evaluate the integrals appearing in Eqs. (3.19), we used the so-called Puso's approximation for the inverse Langevin function \cite{puso1994mechanistic}

\begin{equation}
\pounds^{-1}(\beta)\approx\frac{3\beta}{1-\beta^{3}}
\end{equation}

 \noindent and the resulting integrals were calculated using MATLAB.  The results are expressed in terms of non-dimensional parameters $\overline{\kappa}_{0}=\frac{k^{0}_{r}}{k_{f}}$ and $\overline{\delta}=\frac{\delta}{l}$ satisfying
 
 \begin{equation}
 \overline{\kappa}=\frac{k_{r}}{k_{f}}=\overline{\kappa}_{0}\textrm{exp}\big[\beta \overline{\delta}\big]
 \end{equation}
 
 \noindent In the presented examples, the binding energy between polymer segments is assumed to be much stronger than the adsorption between strands and the filler surface; i.e. $\xi^{(b)}\gg\xi^{(a)}$, unless otherwise is explicitly mentioned. Furthermore, the volume fraction of the transition zone is taken to be equal to the volume fraction of fillers ($\nu_{t}=\nu{f})$.   

\bigskip
  Figure 3.3 shows how the polydispersity affects the steady-state distribution of strands in the transition zone. We calculated $P^{(f)}(j)$ and $P^{(b)}(j)$ for a random network with $\overline{j}=80$ in the bulk. The sharp decrease in  distribution of bound strands indicates a large population of short adsorbed strands. The short strands are expected to act as culprits in causing rupture as they can be easily desorbed by a small macroscopic stretch. Figure 3.4, shows the response of this system to a uniaxial deformation and compares it with filled networks at different $\overline{j}$ values. We chose a range of the $\overline{j}$ values comparable with the simulation results of Svaneborg et al. \cite{svaneborg2005disorder} for unfilled random networks. The effect of polydispersity on the ultimate strength is apparent. Comparatively, the networks with smaller $\overline{j}$ exhibit significantly lower strength. Due to the chosen value of $\xi^{(b)}$, the network degradation is essentially due to the rupture of labile bonds and adhesive failure of the adsorbed strands. To observe the effect of polydispersity more clearly, Figure 3.4 also includes the stress-stretch dependency of a monodisperse network reinforced with the same volume fraction of fillers, where all strands contain 80 statistical segments. The drastic difference between the behavior of monodisperse and polydisperse networks confirms that the network alteration is indeed induced by the structural polydispersity. The deformation-induces desorption is driven by the finite extensibility and non-Gaussian behavior of short strands in the transition zone. The entropic tension developed in shorter strands increases concurrent with macroscopic deformation and leads to desorption of a large population of short strands. In the presented model, the progressive desorption of strands is taken into account by evolution of $j^{(a)}_{min}$ with $\lambda$ (Figure 3.5).

\begin{figure}[H]
	\centering
	\includegraphics[width=0.85\linewidth]{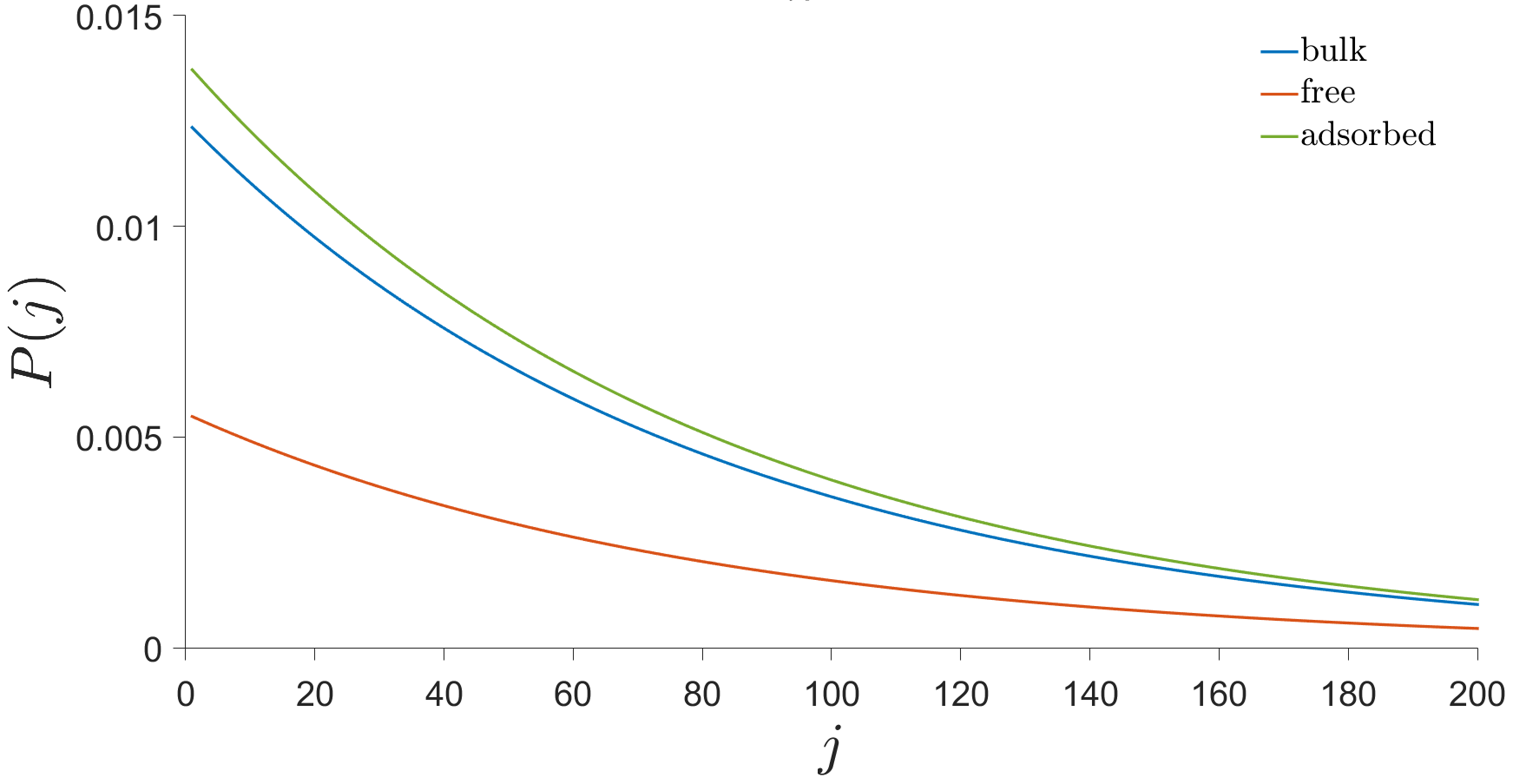}
	\caption{Distribution of strands in transient zone ($\overline{j}=80$, $\nu_{f}=0.1$, $\overline{\kappa}_{0}=2.5$, $\overline{\delta}=0.2$).}
	\label{fig:boat1}
\end{figure}
\begin{figure}[H]
	\centering
	\includegraphics[width=0.85\linewidth]{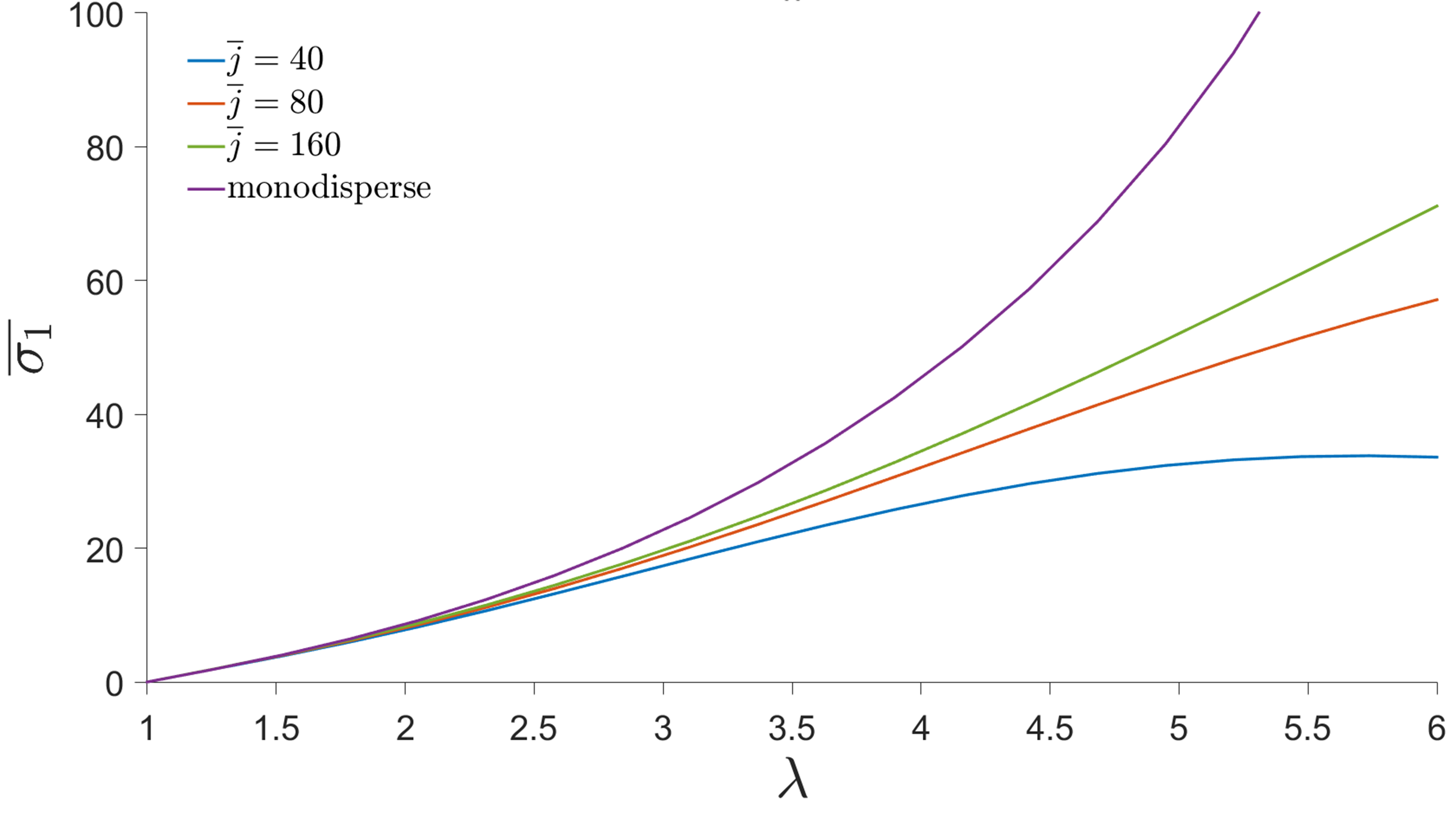}
	\caption{The effect of polydispersity parameter, $\overline{j}$, on the stress behavior of random networks subjected to uniaxial deformation ($\overline{\sigma_{1}}=\sigma_{1}/\mu$, $\nu_{f}=0.1$, $\xi^{(b)}=\xi^{(f)}=0.99$, $\xi^{(a)}=0.2$, $\overline{\kappa}=2.5$, $\overline{\delta}=0.2$).}
	\label{fig:boat1}
\end{figure}

\begin{figure}[H]
	\centering
	\includegraphics[width=0.85\linewidth]{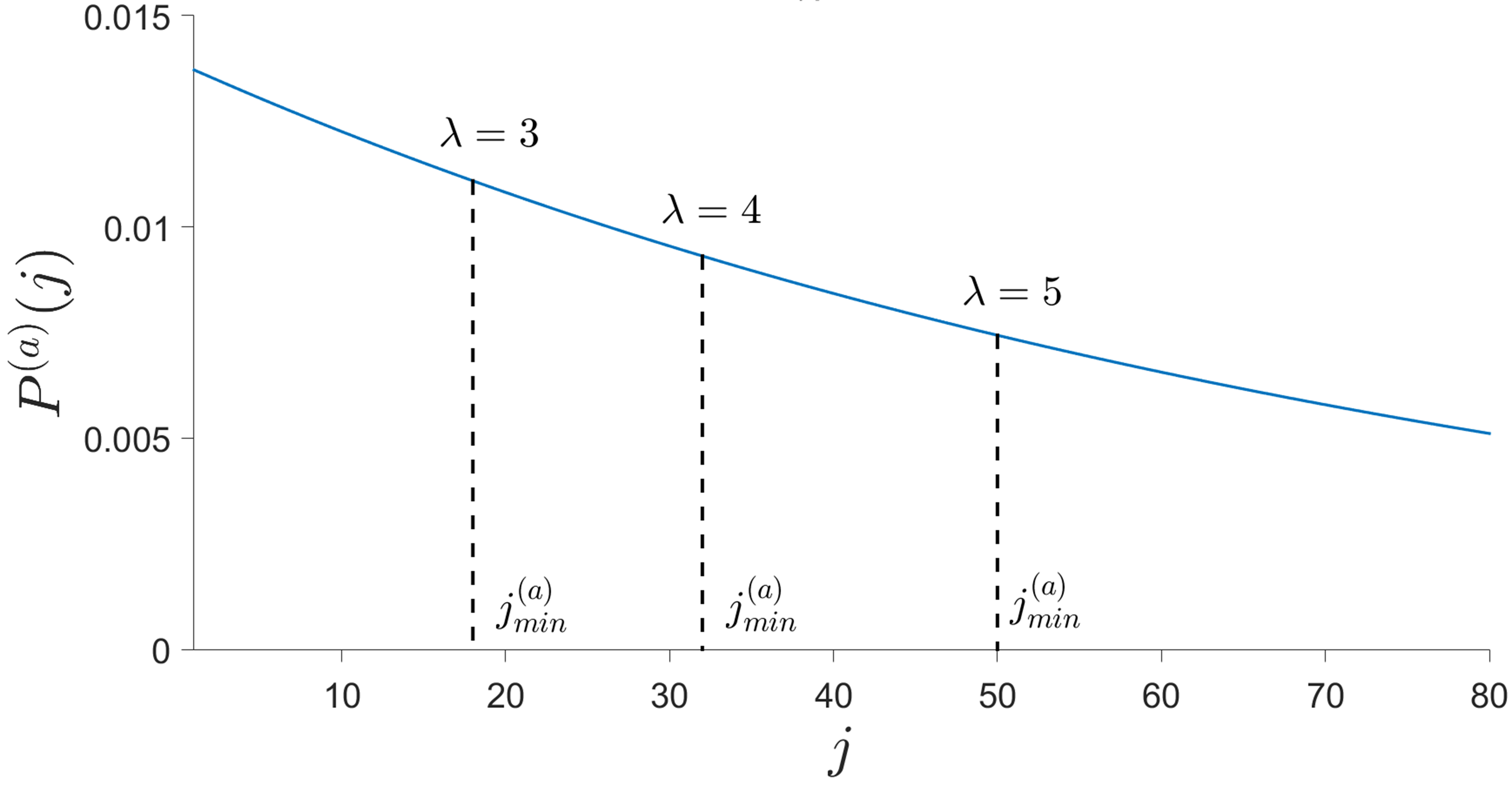}
	\caption{Variation of $j_{min}^{(a)}$ of adsorbed strands during uniaxial deformation($\overline{j}=80$, $\nu_{f}=0.1$, $\xi^{(b)}=\xi^{(f)}=0.99$, $\xi^{(a)}=0.2$, $\overline{\kappa}_{0}=2.5$, $\overline{\delta}=0.2$).}
	\label{fig:boat1}
\end{figure}
\newpage
 Figure 3.6 shows the effect of relative values of strength parameters $\xi^{(a)}$ and $\xi^{(b)}$ on the variation of overall stress in the filled network. If the cohesive energy between strand monomers is strong  (i.e., $\xi^{(b)} \rightarrow 1$) increasing the energetic affinity between strands and fillers leads to enhancement of the ultimate strength of the composite. This is frequently reported in experimental measurements For example, surface modification of carbon black with organic functional groups or fatty acids or increasing the hydrogen content on surface of carbon black are reported to moderately improve the tensile properties of filled SBR \cite{han2006effect,ghosh1997modified,ghosh1999reinforcing,ganguly2005effect}. In case of silica particles, the interaction between fillers and polymer can be enhanced by providing covalent bonds. The surface chemistry of silica particles is different from carbon black, primarily due to superficial hydroxyl and geminal silanol groups\cite{Donnet1994}. Bifunctional organosilanes have been effectively used, particularly in tire industry, to facilitate the covalent bonding between silica particles and rubber molecules\cite{byers1998silane,hashim1998effect}. It has been proposed that if rubber-particle adhesion is stronger than the cohesion of rubber matrix, the damage may occur in the bulk instead of the interface\cite{suzuki2005effects}. Our model, qualitatively predicts such possibility when $\xi^{(a)} > \xi^{(b)}$, in which case the softening in mechanical response is essentially due to the alteration of bulk strands (for example by chain scission). 

\begin{figure}[H]
	\centering
	\includegraphics[width=0.85\linewidth]{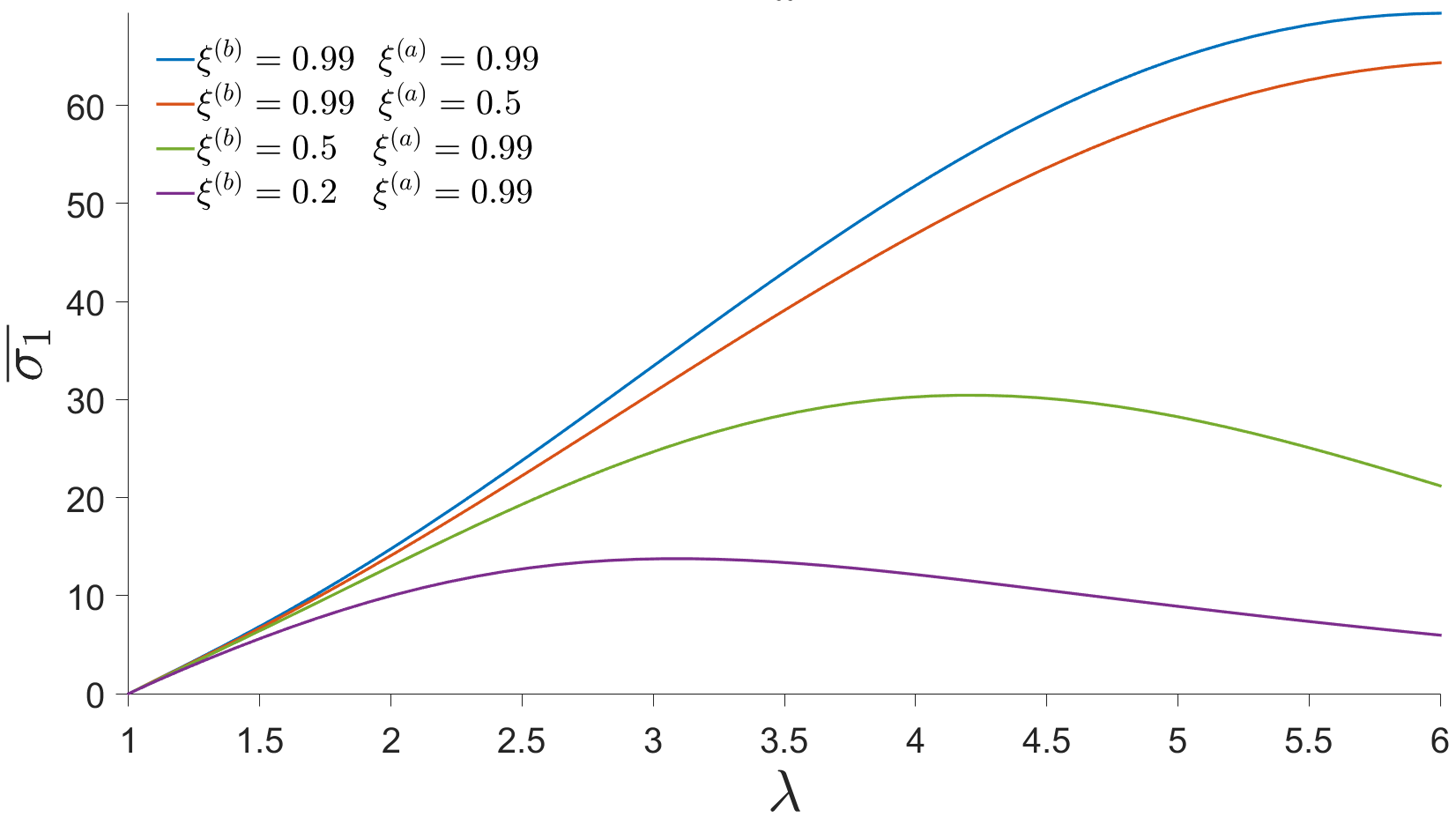}
	\caption{The effect of bond strength parameter, $\xi$, on the stress behavior of random networks subjected to uniaxial deformation ($\overline{\sigma_{1}}=\sigma_{1}/\mu$, $\overline{j}=80$, $\nu_{f}=0.1$, $\overline{\kappa}_{0}=2.5$, $\overline{\delta}=0.2$).}
	\label{fig:boat1}
\end{figure}

\bigskip

\begin{figure}[H]
	\centering
	\includegraphics[width=0.75\linewidth]{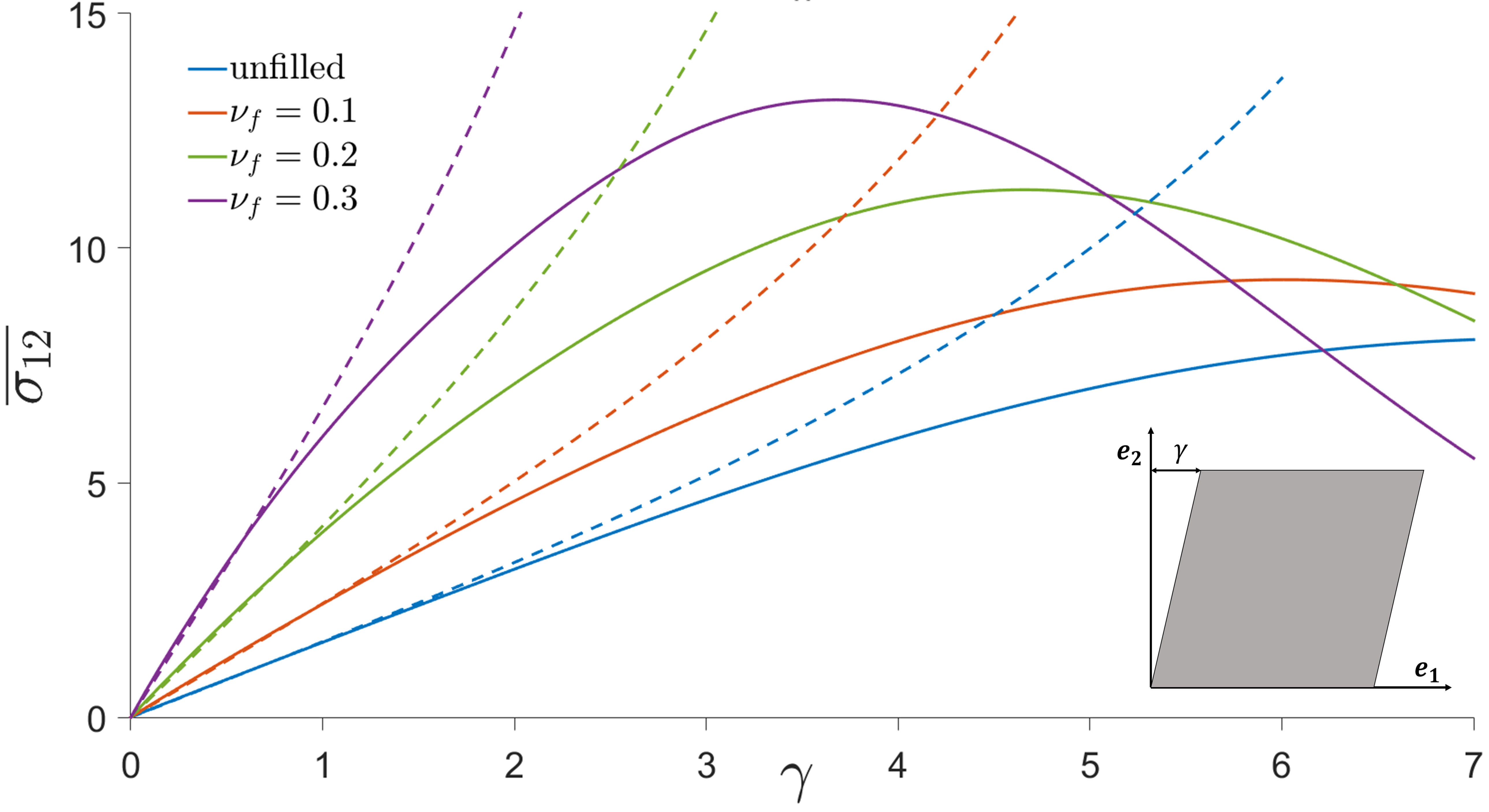}
	\caption{The effect of fillers content, $\nu_{f}$, on the stress behavior of networks subjected to simple shear deformation ($\overline{\sigma_{12}}=\sigma_{12}/\mu$, $\overline{j}=80$, $\xi^{(b)}=\xi^{(f)}=0.99$, $\xi^{(a)}=0.2$, $\overline{\kappa}_{0}=2.5$, $\overline{\delta}=0.2$) Dashed lines show the response of monodisperse networks whose strands are formed from 80 segments at different filler contents.}
	\label{fig:boat1}
\end{figure}

 As expected, the volume fraction of particles markedly affects the mechanical behavior of the network (Figure 3.7). A larger filler content improves the strength but has a compromising effect on the deformation at break. Enhanced strength can be attributed to the amplification of stretch field by hydrodynamic correction factor $X$ and a significant stress production by highly stretched short strands in the transition zone. Increasing deformation is followed by desorption of short strands, which controls the onset of softening and reduces the deformation at break with increasing the filler content. Finally, to examine the predictive ability of the proposed model, we compared the model predictions with a collection of experimental data represented by Meissner and Mat\v{e}jka \cite{meissner2001description}. They collected a series of experimental data presenting the stress-stretch dependency of unfilled and filler rubbers subjected to uniaxial deformation. We chose the result for SBR reinforced with different fractions of carbon blacks. The curves are shown in the so-called Mooney-Rivlin coordinates. The model parameters are obtained by fitting to the experimental data of neat SBR. The calibrated model was used to predict the response of filled networks with 20\% and 30\% carbon black, without changing any other model parameter.   
\begin{figure}[H]
	\centering
	\includegraphics[width=0.75\linewidth]{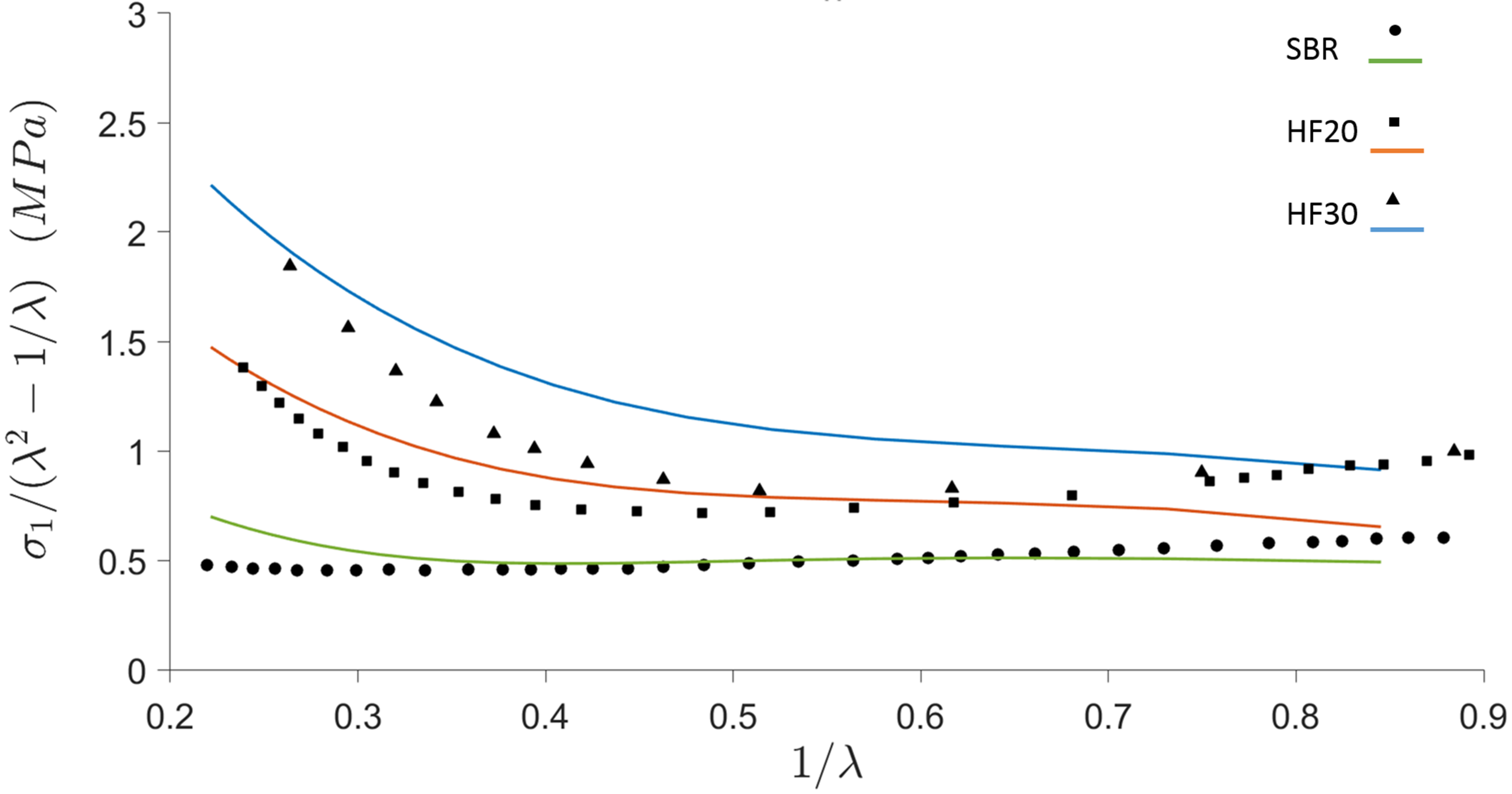}
	\caption{The comparison between proposed model (solid lines) and the experimental data (symbols) presented by Meissner and Mat\v{e}jka \cite{meissner2001description}($\mu=0.2875$, $\overline{j}=160$, $\xi^{(b)}=\xi^{(f)}=0.99$, $\xi^{(a)}=0.1$, $\overline{\kappa}_{0}=2.5$, $\overline{\delta}=0.2$).}
	\label{fig:boat1}
\end{figure}

\section{Concluding Remarks}

 In this contribution, we proposed a new mechanism that is sufficient to cause marked alteration in the structure of filled polymer networks and can potentially influence the damage initiation and control the softening of filled elastomers in response to deformation. The proposed theory is based on the assumption that internal structure of vulcanizates is random and the distribution of strand length is polydisperse in nature. After compounding filler particles into a rubber matrix, each particle interacts with several polymer strands each of different length. This energetic interaction forms a myriad of reversible attachments between strands and particle surface. It was proposed that the alteration in mechanical behavior results from desorption of these physical bonds. During deformation shorter adsorbed strands reach their maximum elongation first and then snap. The cohesive failure of short and highly elongated strands continues concurrently with applied deformation. The softening can thus be attributed to a significant reduction in density of adsorbed strands and progressive loss of friction between the filler and rubber matrix. Although predicted for static loading, this scenario may well explain some of the main characteristic features of the Payne effect in filled rubbers subjected to dynamic deformations. Namely, it describes the direct correlation between storage modulus with the filler content at small deformations, the drop in storage modulus with increasing strain, and shifting the onset of nonlinearity to lower deformations with increasing the filler content. 

\bigskip
 The model is built upon multiple simplifying assumptions. Probably the most tenuous assumption here is the affine motion of all strands. Furthermore, the Morse potential is only a naive representation of bonding along the backbone and adhesion to the filler surface. Even though the presented model is a too simple of a picture, it provides compelling numerical evidences to support the effects of polydispersity on mechanical performance of filled vulcanizates. The main challenge, however, is to directly measure the strand length distribution in randomly crosslinked networks. Long ago, Gehman\cite{gehmanmolecular} referred to the distribution of network strands as a \textquotedblleft mental concept\textquotedblright that cannot be directly measured. Precise observations are eluded primarily due to the insolubility of crosslinked networks. Indirect measurements such as relaxometry of stressed networks or measurement of post-swelling pressure only provided limited information about microstructural irregularities in polymer networks. Light scattering and small-angle X-ray and neutron scattering techniques, however, have confirmed spatial fluctuation in crosslink density in networks on scales of 1-100 nm. Further developments along this line may eventually help us obtain a correct estimate of network chain distribution, a challenge that still exists today. 

\chapter{Summary}

 In this study, the effect of randomness in internal structure of both unfilled and filled polymer networks on their mechanical behavior was studied. In the first study, a predictive model was formulated in order to model the elasticity and initiation of bulk damage in unfilled polydisperse networks subjected to a finite deformation. The polydispersity was represented by an exponential distribution function formulated for polymer vulcanizates by Bueche \cite{bueche} and Watson \cite{watson1953chain,watson1954chain}. It was shown that in a polydisperse network, the ultimate mechanical strength directly correlates with the population of short strands. The damage was initiated by breaking of shorter strands that experience stronger entropic tension due to Langevin effect. The progressive damage of strands continues with deformation and determines the ultimate mechanical strength of the network. The theory was also used to predict the history-dependent damage of random networks under slow cyclic loading. This led to the conclusion that polydispersity may play a detrimental role in fatigue life of polymer vulcanizates.\par

\bigskip
 The study was extended to examine the effect of network polydispersity on damage initiation and strength of the  networks reinforced with filler particles. The  network was assumed to be reversibly interacting with the surface of filler by forming  strong physical bonds. Kinetics of polymer adsorption was taken into account by a set of first-order kinetic equations within a transition zone surrounding each particle. The model was able to predict certain characteristic features of mechanical behavior of filled rubbers. Namely, it describes the correlation between stiffness with the filler content at small deformations, the drop in network stiffness with increasing stretch, and shifting the onset of nonlinearity to the lower deformation with increasing the filler content. These conclusions lend credence to the hypothesis that polydispersity may also control the mechanism of damage initiation in filled rubber vulcanizates.

\references{
	\bibliography{mythesisbib}
}

\appendix           

\chapter{Appendix A}

\noindent Eq. (2.20) can be written as

\begin{multline} 
	\boldsymbol{\sigma}=\mu \sum_{k=1}^{3} \lambda_{k}(\boldsymbol{n}^{(k)}\otimes  \boldsymbol{n}^{(k)})  \   \Big(\int\limits_{0}^{2\pi}\int\limits_{0}^{\pi} \int\limits_{j_{min}(\lambda)}^{\infty} \ P(j) \  \frac{\partial}{\partial \lambda_{k}}  w(\lambda , j) \ \sin\theta_{0} dj d\theta_{0} d\phi_{0}  \\
	- \int\limits_{0}^{2\pi}\int\limits_{0}^{\pi}  \frac{\partial j_{min}(\lambda)}{\partial \lambda_{k}} \ P(j_{min}) \   w(\lambda,j_{min}) \ \sin\theta_{0} dj d\theta_{0} d\phi_{0}  \Big)
\end{multline}

\noindent where

\begin{equation} 
	\frac{\partial w(\lambda_{r},j)}{\lambda_{k}}=\frac{\partial \lambda}{\partial \lambda_{k}} \frac{\partial}{\partial \lambda}  w(\lambda)
\end{equation}

\noindent Using Eq. (2.12), we obtain

\begin{equation} 
	\frac{\partial \lambda}{\partial \lambda_{1}}=\frac{\lambda_{1}}{\lambda}\sin^{2}\theta_{0} \cos^{2}\phi_{0}
\end{equation}

\begin{equation} 
	\frac{\partial \lambda}{\partial \lambda_{2}}=\frac{\lambda_{2}}{\lambda} \sin^{2}\theta_{0} \sin^{2}\phi_{0}
\end{equation}

\begin{equation} 
	\frac{\partial \lambda}{\partial \lambda_{3}}=\frac{\lambda_{3}}{\lambda} \cos^{2}\theta_{0}
\end{equation}

\noindent Taking partial derivative of the strain energy of a single strand with respect to the macroscopic stretch yields 

\begin{equation} 
	\frac{\partial}{\partial \lambda}  w(\lambda) =  \beta+\lambda\frac{\partial \beta}{\partial \lambda} +\frac{1}{\beta}\frac{\partial \beta}{\partial \lambda}-\frac{\partial \beta}{\partial \lambda}\coth(\beta)
\end{equation}

\noindent The inverse Langevin function can be approximated in different ways \cite{jedynak2015approximation}. Here, we used the so called Puso's approximation \cite{puso1994mechanistic}

\begin{equation} 
	\beta=\pounds^{-1}(y)\approx\frac{3y}{1-y^{3}}
\end{equation}

\noindent which leads to

\begin{equation} 
	\frac{\partial}{\partial y}\pounds^{-1}(y)\approx\frac{3+6y^3}{(1-y^{3})^{2}}
\end{equation}

\end{document}